\documentclass[a4paper,12pt]{article}
\usepackage[leqno]{amsmath}
\usepackage{amssymb}
\usepackage{amsthm}
\usepackage[mathscr]{eucal}
\usepackage{diagrams}
\usepackage{url}
\usepackage{hyperref}

\IfFileExists{Ueur.fd}{\input{Ueur.fd}}{\input{ueur.fd}}
\DeclareMathAlphabet{\eurm}{U}{eur}{m}{n}
\DeclareMathAlphabet{\eubf}{U}{eur}{b}{n}
\DeclareFontFamily{U}{UWCyr}{}
\DeclareFontShape{U}{UWCyr}{m}{n}{%
  <5> <6> <7> <8> <9>
  <10> <10.95> <12> <14.4> <17.28> <20.74> <24.88> wncyr10
  }{}
\DeclareFontShape{U}{UWCyr}{m}{it}{%
  <5> <6> <7> <8> <9>
  <10> <10.95> <12> <14.4> <17.28> <20.74> <24.88> wncyi10
  }{}
\DeclareFontShape{U}{UWCyr}{m}{sc}{%
  <5> <6> <7> <8> <9>
  <10> <10.95> <12> <14.4> <17.28> <20.74> <24.88> wncysc10
  }{}
\DeclareFontShape{U}{UWCyr}{b}{n}{%
  <5> <6> <7> <8> <9>
  <10> <10.95> <12> <14.4> <17.28> <20.74> <24.88> wncyb10
  }{}
\DeclareMathAlphabet{\cyrm}{U}{UWCyr}{m}{n}
\DeclareMathAlphabet{\cyit}{U}{UWCyr}{m}{it}
\DeclareMathAlphabet{\cysc}{U}{UWCyr}{m}{sc}
\DeclareMathAlphabet{\cybf}{U}{UWCyr}{b}{n}
\newtheoremstyle
{MyThm}
{10pt}
{10pt}
{\itshape}
{\parindent}
{\bfseries}
{.}
{.5em}
{}
\swapnumbers
\theoremstyle{MyThm}
\newcounter{assump}
\newtheorem{Assumption}{Assumption}[assump]

\newcounter{postul}
\newtheorem{Postulate}{Postulate}[postul]

\newtheorem{Statement}{Statement}[section]
\newtheorem{Caution}[Statement]{Caution}
\newtheorem{Convention}[Statement]{Convention}
\newtheorem{Corollary}[Statement]{Corollary}
\newtheorem{Definition}[Statement]{Definition}
\newtheorem{Example}[Statement]{Example}
\newtheorem{Exercise}[Statement]{Exercise}
\newtheorem{Interpretation}[Statement]{Interpretation}
\newtheorem{Lemma}[Statement]{Lemma}
\newtheorem{Notation}[Statement]{Notation}
\newtheorem{Note}[Statement]{Note}
\newtheorem{Problem}[Statement]{Problem}
\newtheorem{Proposition}[Statement]{Proposition}
\newtheorem{Remark}[Statement]{Remark}
\newtheorem{Sequence}[Statement]{Sequence}
\newtheorem{Theorem}[Statement]{Theorem}
\newcommand{\bAs}{\begin{Assumption}\em}
\newcommand{\eAs}{\end{Assumption}}
\newcommand{\bCa}{\begin{Caution}\em}
\newcommand{\eCa}{\end{Caution}}
\newcommand{\bCr}{\begin{Corollary}\em}
\newcommand{\eCr}{\end{Corollary}}
\newcommand{\bCv}{\begin{Convention}\em}
\newcommand{\eCv}{\end{Convention}}
\newcommand{\bDf}{\begin{Definition}\em}
\newcommand{\eDf}{\end{Definition}}
\newcommand{\bDr}{\begin{Exercise}\em}
\newcommand{\eDr}{\end{Exercise}}
\newcommand{\bEx}{\begin{Example}\em}
\newcommand{\eEx}{\end{Example}}
\newcommand{\bIn}{\begin{Interpretation}\em}
\newcommand{\eIn}{\end{Interpretation}}
\newcommand{\bLm}{\begin{Lemma}\em}
\newcommand{\eLm}{\end{Lemma}}
\newcommand{\bNo}{\begin{Notation}\em}
\newcommand{\eNo}{\end{Notation}}
\newcommand{\bNt}{\begin{Note}\em}
\newcommand{\eNt}{\end{Note}}

\newcommand{\bPb}{\begin{Problem}\em}
\newcommand{\ePb}{\end{Problem}}
\newcommand{\bPf}{\begin{proof}[\noindent\indent{\sc Proof}]}
\newcommand{\ePf}{\renewcommand{\qedsymbol}{}\end{proof}}
\newcommand{\bpf}{\bfz\bPf}
\newcommand{\epf}{\ePf\efz}
\newcommand{\bPr}{\begin{Proposition}\em}
\newcommand{\ePr}{\end{Proposition}}
\newcommand{\bPs}{\begin{Postulate}\em}
\newcommand{\ePs}{\end{Postulate}}
\newcommand{\bRm}{\begin{Remark}\em}
\newcommand{\eRm}{\end{Remark}}

\newcommand{\bSq}{\begin{Sequence}\em}
\newcommand{\eSq}{\end{Sequence}}
\newcommand{\bSt}{\begin{Statement}\em}
\newcommand{\eSt}{\end{Statement}}
\newcommand{\bTh}{\begin{Theorem}}
\newcommand{\eTh}{\end{Theorem}}

\newcommand{\bEq}{\begin{equation}}
\newcommand{\eEq}{\end{equation}}
\newcommand{\beq}{\begin{equation*}}
\newcommand{\eeq}{\end{equation*}}
\newcommand{\bal}{\begin{align*}}
\newcommand{\bAl}{\begin{align}}
\newcommand{\bat}{\begin{alignat*}}
\newcommand{\bAt}{\begin{alignat}}
\newcommand{\bml}{\begin{multline*}}
\newcommand{\bMl}{\begin{multline}}
\newcommand{\bgt}{\begin{gather*}}
\newcommand{\bGt}{\begin{gather}}
\newcommand{\bCd}{\bEq\begin{CD}}
\newcommand{\eCd}{\end{CD}\eEq}
\newcommand{\bcd}{\beq\begin{CD}}
\newcommand{\ecd}{\end{CD}\eeq}
\newcommand{\bdg}{\beq\begin{diagram}}
\newcommand{\edg}{\end{diagram}\eeq}
\newcommand{\bDg}{\bEq\begin{diagram}}
\newcommand{\eDg}{\end{diagram}\eEq}
\newcommand{\bmt}{\left(\begin{matrix}}
\newcommand{\emt}{\end{matrix}\right)}
\newcommand{\bcn}{\begin{center}}
\newcommand{\ecn}{\end{center}}
\newcommand{\ben}{\begin{enumerate}}
\newcommand{\een}{\end{enumerate}}
\newcommand{\btb}{\begin{tabbing}}
\newcommand{\etb}{\end{tabbing}}
\newcommand{\bsm}{\begin{quotation}\small}
\newcommand{\esm}{\end{quotation}}
\newcommand{\bfz}{\begin{footnotesize}}
\newcommand{\efz}{\end{footnotesize}}
\newcommand{\bsz}{\begin{scriptsize}}
\newcommand{\esz}{\end{scriptsize}}

\newcommand{\fz}{\footnotesize}

\newcommand{\bsb}
{\vspace{-0.8cm}
\begin{alignat*}{2}
& \qquad\qquad\qquad\qquad\qquad\qquad\qquad\qquad\qquad\qquad
&&\qquad\qquad\qquad\qquad\qquad\qquad\qquad\qquad\qquad
\\}

\newcommand{\Rn}{\text{I\!R}}

\newcommand{\h}{\hbar}

\newcommand{\der}{\partial}

\newcommand{\nab}{\nabla}

\newcommand{\Upa}{^{\uparrow}{}}

\newcommand{\Ele}{^\mathfrak{e}{}}
\newcommand{\Elg}{^\mathfrak{g}{}}

\newcommand{\Fla}{^{\flat}{}}
\newcommand{\Sha}{^{\sharp}{}}

\newcommand{\prl}{_{\|}{}}

\newcommand{\per}{_{\perp}{}}

\newcommand{\lang}{\langle}
\newcommand{\rang}{\rangle}

\newcommand{\mto}{\mapsto}

\newcommand{\com}{\circ}
\newcommand{\comm}{\!\circ\!}
\newcommand{\car}{\times}
\newcommand{\ten}{\otimes}

\newcommand{\wed}{\wedge}

\DeclareMathOperator{\con}{\lrcorner}

\newcommand{\nid}{\not\equiv}
\DeclareMathOperator{\byd}{\,{\raisebox{.092ex}{\rm :}{\rm =}}\,}

\newcommand{\udrs}[1]{\underset{#1}{\oplus}}

\newcommand{\uten}[1]{\underset{#1}{\otimes}}

\newcommand{\db}[1]{{\,{#1}\!{#1}\,}}

\newcommand{\fr}[2]{\frac{#1}{#2}\,}
\newcommand{\tfr}[2]{\tfrac{#1}{#2}\,}

\newcommand{\col}[3]{_{#1}{}^{#2}{}_{#3}}

\newcommand{\Ga}[2]{_{#1}{}^{#2}_0}

\newcommand{\END}{{\,\text{\footnotesize\qedsymbol}}}
\newcommand{\QED}{{\,\text{\rm{\footnotesize QED}}}}

\newcommand{\ssep}[1]{{\qquad\text{\rm{#1}}\qquad}}

\newcommand{\bi}{\bibitem}
\newcommand{\au}[1]{{\sc#1}:}
\newcommand{\tp}[1]{\emph{#1},}

\newcommand{\pu}[1]{#1.}
\newcommand{\ar}[1]{\url{http://arXiv.org/abs/#1}}

\DeclareMathOperator{\id}{{{id}}}

\DeclareMathOperator{\spec}{{{spec}}}

\newcommand{\emp}{\emph}

\newcommand{\f}[1]{{\boldsymbol{#1}}}

\newcommand{\ba}[1]{{{\bar{#1}}}}

\newcommand{\ha}[1]{{\hat{#1}}}
\newcommand{\wha}[1]{{\widehat{#1}}}

\newcommand{\dt}[1]{{\dot{#1}}}

\newcommand{\br}[1]{{\breve{#1}}{}}

\newcommand{\bma}{\left(\begin{matrix}}
\newcommand{\ema}{\end{matrix}\right)}

\newcommand{\C}[1]{{\mathcal{#1}}}

\newcommand{\M}[1]{{\mathscr{#1}}}

\newcommand{\B}[1]{{\mathbb{#1}}}

\newcommand{\baB}[1]{{\bar{{\mathbb{#1}}}}}

\newcommand{\K}[1]{{\cyrm{#1}}}

\newcommand{\alp}{\alpha}

\newcommand{\gam}{\gamma}
\newcommand{\del}{\delta}

\newcommand{\tht}{\theta}

\newcommand{\kap}{\kappa}

\newcommand{\lam}{\lambda}

\newcommand{\sig}{\sigma}

\newcommand{\ome}{\omega}
\newcommand{\Gam}{\Gamma}

\newcommand{\Lam}{\Lambda}

\newcommand{\Sig}{\Sigma}

\newcommand{\Ome}{\Omega}

\DeclareMathOperator{\Ad}{Ad}
\DeclareMathOperator{\CoAd}{CoAd}

\DeclareMathOperator{\reg}{{{reg}}}

\newcommand{\ual}{\underline{\alpha}}
\newcommand{\odx}{\overline{d}}
\newcommand{\brd}[1]{\br\del^{#1}_0}
\newcommand{\cprime}{\/{\mathsurround=0pt$'$}}

\title{\textbf{On the characterization\\ of infinitesimal symmetries\\
    of the relativistic phase space}}
\author{
\bf 
Josef Jany\v ska$^1$, Raffaele Vitolo$^2$
\bigskip
\\
\fz $^1$Department of Mathematics and Statistics, Masaryk University
\\
\fz Kotl\'a\v rsk\'a 2, 611 37 Brno, Czech Republic
\\
\fz email: \texttt{janyska@math.muni.cz} \medskip
\\
\fz $^2$Department of Mathematics and Physics ``E. De Giorgi'', University of
Lecce
\\
\fz Via per Arnesano, 73100 Lecce, Italy
\\
\fz email: {\tt raffaele.vitolo@unisalento.it}
}

\date{
\framebox{
\begin{minipage}{8cm}
\begin{center}
\footnotesize
Published in \emph{Journal of Physics A: Math. Theor.},
\\
Vol.\ 45 (2012) 485205 (28pp).
\end{center}
\end{minipage}
}
}

\begin{document}
\maketitle
\begin{abstract}
  The phase space of relativistic particle mechanics is defined as the 1st jet
  space of motions regarded as timelike 1--dimensional submanifolds of
  spacetime.  A Lorentzian metric and an electromagnetic 2-form define
  naturally on the odd-dimensional phase space a generalized contact
  structure. In the paper infinitesimal symmetries of the phase structures are
  characterized. More precisely, it is proved that all phase infinitesimal
  symmetries are special Hamiltonian lifts of distinguished conserved
  quantities on the phase space. It is proved that generators of infinitesimal
  symmetries constitute a Lie algebra with respect to a special bracket.  A
  momentum map for groups of symmetries of the geometric structures is
  provided.


\end{abstract}
\section*{Introduction}

Relativistic mechanics of particles with mass is usually formulated, in a
differential geometric context, as a theory on the tangent bundle of spacetime
endowed with the symplectic form induced by the pseudo-Riemannian metric.
Anyway, this formulation has the drawback of presenting a degenerate
Lagrangian.  The degeneracy is due to the fact that the Lagrangian is invariant
with respect to the affine reparametrization of curves, which acts as a gauge
group of the theory \cite{FGT99,HenTei92}.  As a way to solve it, some
authors tried to formulate mechanics on `trajectories', or non-parametrized
curves \cite{NiTo83,Sar98}. None of them, however, present a complete model for
the relativistic mechanics.

Recently, a formulation of relativistic mechanics based on jets of submanifolds
and contact forms has been presented \cite{JanMod95,JanMod08}. Here the tangent
bundle is replaced by the first jet space of time-like curves as the phase
space of the theory. Time-like curves can be regarded as distinguished
$1$-dimensional submanifolds: indeed, the phase space is a subspace of the
first jet of $1$-dimensional submanifolds of spacetime.

The literature on jets of submanifolds (also known as `contact elements',
`differentiable elements', `first-order caps', `extended jets' etc., see
\cite{ALV91,KMS93,Olv92} and references therein) is less spread with respect to
the literature on jets of fibrings.  Jets of $1$-dimensional submanifolds of
order $r$ are equivalence classes of $1$-dimensional submanifolds having a
contact of order $r$.  Clearly, two curves have a contact if and only if their
tangent space is the same; hence this notion is independent of the
parametrization.  In particular, the phase space of relativistic mechanics is
the set of first jets of time-like curves, and in the case of vanishing
electromagnetic field, the equation of motion is the equation for
unparametrized geodesics, or geodesics as 1-dimensional
submanifolds. Reparametrization-invariant theories are quite common
(\emph{e.g.}, string theories, \cite{FGT99}) and it is always possible to get
rid of reparametrization invariance without any gauge-fixing by using jets of
submanifolds, which can be regarded as ordinary jets factored by the
reparametrization group \cite{KMS93}.

In \cite{JanMod95,JanMod08,JanMod09} a structure generalizing contact and
cosymplectic structures (in the sense of
\cite{Albert89,dLS93,deLTuy96,LiMa87,Lich78}) naturally induced from the
gravitational and the electromagnetic fields has been introduced on the first
jet space of time-like curves. Such a structure is a special case of
\emph{generalized contact structure}, \cite{PooWad11, Wad12}, that we call
\emph{almost-cosymplectic-contact structure}. The structure has a
one-dimensional kernel (Reeb vector field) whose integral curves are the
trajectories of the system. A non-degenerate Lagrangian formulation of the
theory was developed in \cite{ManVit05} using the variational calculus on jets
of submanifolds \cite{Ded78,ManVit08,Vin78}. Indeed, our model is a universal
mass shell, \emph{i.e.} it is isomorphic to a mass shell \cite{HenTei92} for
each choice of a particle mass. In Dirac's languange, we work on a primary
constraint which is naturally endowed with a generalized contact structure. A
Hamiltonian formulation is straightforward in this context.

Then, a quantum theory for a scalar particle on the above background has been
formulated by analogy with the geometric quantization of classical mechanical
systems \cite{Vit00}, and this leads to a covariant Klein--Gordon equation
\cite{Jan98}.

In this paper we study infinitesimal symmetries of the geometrical structures
which are naturally given on the phase space by a pseudo-Riemannian metric $g$
and an electromagnetic field $F$.

The study of such infinitesimal symmetries dates back to \cite{Iwa76}, where
the phase space was taken to be the (odd-dimensional) pseudosphere subbundle of
the tangent bundle of spacetime. The geometrical structure of such phase space
is obtained as the restriction of the canonical symplectic structure given on
the tangent bundle of the spacetime by $g$ and $F$. It can be easily realized
that the $1$-jet of time-like submanifolds and the pseudosphere subbundle are
isomorphic, but the advantage of jet space over pseudosphere is that, while the
pseudosphere is defined by an equation, there exists on the jet space a natural
choice of coordinates induced by spacetime coordinates. This makes the
computations much easier. The main result in \cite{Iwa76} is
that the tangent lift of a vector field $X$ on $\f E$ is an infinitesimal
symmetry of the phase structure if and only if $X$ is an infinitesimal symmetry
of $g$ and $F$, i.e.  $L_X g = 0$ and $L_X F = 0$. Moreover, conserved
quantities related with the above symmetries are described.

In our approach to the phase space we obtain results which are ``parallel'' to
results for Galilean spacetime in
\cite{JanModSal11,ModSalTol06,SalVit00}. Namely, we prove that \emph{all}
projectable infinitesimal symmetries of the generalized contact structure are
holonomic lifts to the phase space of infinitesimal symmetries of $g$ and $F$.
Here, ``projectable'' means that we consider infinitesimal symmetries that
project onto vector fields on spacetime, and ``holonomic lift'' means the lift
of a vector field to the first jet space.

Moreover, we prove that all such infinitesimal symmetries can be obtained as
special Hamiltonian lifts (\emph{i.e.}, lifts defined through the generalized
contact structure) of distinguished functions on the phase space, which we call
special phase functions. Special phase functions are uniquely determined by a
spacetime vector field and a function on spacetime, and have the form defined
by eq.~\ref{Special function}. In the case when they generate an infinitesimal
symmetry of the generalized contact structure, the vector field is a symmetry
of $g$ and $F$ and the special phase function is a conserved quantity.  Another
new feature of our approach is that the sheaf of generators of infinitesimal
symmetries constitute a Lie algebra with respect to the special bracket
(\emph{i.e.}, a bracket defined through the generalized contact structure) of
special phase functions.

The above infinitesimal symmetries turn out to be Noether symmetries of the
Lagrangian, hence they are symmetries of the Euler--Lagrange morphism.  This
implies that they are point symmetries of the Euler--Lagrange equation, altough
they do not exhaust that class, see below.

Finally, by analogy with the standard momentum map for Hamiltonian systems
(see, for example, \cite{OrRa04}), we provide an equivariant momentum map for
the generalized contact structure. This will be the starting point for a
Marsden--Weinstein type reduction of the phase space, in the spirit of the
general theory exposed in \cite{Albert89,CC02,dLS93}. As far as we know, this
has never been attempted in relativistic mechanics, and will be the subject of
future work.

As a last comment, we are aware of the fact that the class of symmetries that
we considered in this paper can be substantially enlarged. We do not consider
generalizations to the computation of point or generalized symmetries of the
equation of motion since they would not be symmetries of the geometric
structures on the phase space like the generalized contact structure.  The most
important generalization is obtained by dropping the projectability hypothesis
and dealing with vector fields depending on the phase space in an essential
way. Non-projectable symmetries of this type are called higher
symmetries \cite{Many99}, generalized symmetries \cite{Olv92} or hidden
symmetries (see, for example, \cite{San12}) because they are not related to any
symmetry of spacetime. Non-projectable symmetries of our generalized contact
structure would be Noether symmetries of the equation of motion, like proper
affine or projective symmetries \cite{Hall}, and would
correspond to conserved quantities which can be higher degree polynomials in
velocities. In this case, coefficients have tensor character and constitute the
so-called Killing tensors. Killing-Yano tensors also play an analogous role in
defining hidden symmetries.

Our plan was to focus our attention to symmetries in the context of the
generalized contact geometry of general relativistic spacetime. On the other
hand, as a future research plan it would be interesting to characterize
non-projectable symmetries of the geometric structures on the phase space. At
the present moment, we cannot predict what are the results of the paper that
could be extended to more general symmetries. In particular, we do not know if
it would be possible to define a momentum map, or to associate a wider class of
special phase functions and a Lie bracket between them. See also the discussion
in Section~\ref{sec:perspectives}.

\medskip

\textbf{Acknowledgements.} This work has been partially supported by the
grant GA \v CR 201/09/0981, by the Istituto Nazionale di Alta Matematica
and the Dipartimento di Matematica e Fisica ``E. De Giorgi''. We thank Marco
Modugno for many useful discussions. We also thank the anonymous referees for
many useful remarks that helped us to improve the presentation of our results.
\section{Preliminaries}
\label{Preliminaries}

We assume manifolds and maps to be $\mathcal{C}^{\infty}$.

\textbf{Unit spaces.}  The theory of \emph{unit space} has been developed in
\cite{JanModVit10} in order to make explicit the independence of classical and
quantum mechanics from the choice of unit of measurements. Unit spaces have the
same algebraic structure as $\Rn_+$, but no natural basis. We assume the
($1$-dimensional) unit spaces $\B T$ (space of \emph{time intervals}), $\B L$
(space of \emph{lengths}) and $\B M$ (space of \emph{masses}). We set $\B
T^{-1} \equiv \B T^*$, and analogously for $\B L$, $\B M$.  Tensor fields
appearing in the theory will be usually \emph{scaled}, \emph{i.e.}, they will
take values in unit spaces according to their physical interpretation. For
example, the metric will take values in the space of area units $\B L^2$.

We assume the following constant elements: the \emph{light velocity} $c\in\B
T^{-1}\otimes{\B L}$ and the \emph{Planck's constant} $\hbar\in \B
T^{-1}\otimes\B L^2\otimes\B M$. Moreover, we say a \emph{charge} to be an
element $q\in\B T^{-1}\otimes\B L^{3/2}\otimes\B M^{1/2}\otimes\Rn$.

We will assume coordinates to be dimensionless (\emph{i.e.}, real valued).

\medskip
\textbf{Generalized Lie derivatives.}  We will use a general approach to Lie
derivatives \cite[p. 376]{KMS93}.  If $f:\f M \to \f N$ is a mapping, $X$ a
vector field on $\f M$ and $Y$ a vector field on $\f N$ then we define Lie
derivative of $f$ with respect to the pair $(X,Y)$ as the mapping
$$
L_{(X,Y)} f = Tf \com X - Y \com f : \f M \to T\f N\,.
$$
This definition includes the standard definition of Lie derivative as a special
case. See \cite[p. 59]{KMS93} for more details.

\medskip
\textbf{Infinitesimal symmetries of geometrical structures of odd dimensional
  manifolds.} Let $\f M$ be a $(2n+1)$-dimensional manifold.  A \emph{pre
  cosymplectic (regular) structure (pair)} on $\f M$ is given by a 1-form
$\ome$ and a 2-form $\Ome$ such that $\ome\wed \Ome^n \not\equiv 0$.  A
\emph{contravariant (regular) structure (pair)} $(E,\Lam)$ is given by a vector
field $E$ and a 2-vector $\Lam$ such that $E\wed \Lam^n \not\equiv 0$. We
denote by $\Ome\Fla:T\f M\to T^*\f M$ and $\Lam\Sha:T^*\f M\to T\f M$ the
corresponding "musical" morphisms.  By \cite{Lich78} if $(\ome,\Ome)$ is a pre
cosymplectic pair then there exists a unique regular pair $(E,\Lam)$ such that
\begin{equation}\label{Eq: 2.1}
(\Ome\Fla_{|\text{im}\, \Lam\Sha})^{-1}
= \Lam\Sha_{|\text{im}\, \Ome\Fla} \,,
\quad i_E\ome =1\,,
\quad i_E\Ome =0\,,
\quad i_\ome\Lam = 0\,.
\end{equation}
On the other hand for any regular pair $(E,\Lam)$ there exists a unique
(regular) pair $(\ome,\Ome)$ satisfying the above identities.  The pairs
$(\ome,\Ome)$ and $(E,\Lam)$ satisfying the above identities are said to be
mutually \emph{dual}.  The vector field $E$ is usually called the \emph{Reeb
  vector field} of $(\ome,\Ome)$. A vector field $X$ on $\f M$ is said to be
$\omega$-vertical if and only if $i_X\ome = 0$ and $E$-horizontal if and only
if $i_X\Omega = 0$. Any vector field can be decomposed in a unique way as the
sum of $E$-horizontal and $\omega$-vertical parts.  Similarly a 1-form $\alpha$
on $\f M$ is said to be $E$-vertical if and only if $i_\alpha E = 0$ and
$\omega$-horizontal if and only if $i_\alpha \Lambda = 0$. Any 1-form can be
decomposed in a unique way as the sum of $\omega$-horizontal and $E$-vertical
parts.

Let $(\ome,\Ome)$ and $(E,\Lambda)$ be mutually dual regular structures on  $\f
M$. An \emph{infinitesimal symmetry}  of the structure $(\ome,\Ome)$ is a vector
field $ X $ on $\f M $ such that $L_X\omega = 0$, $L_X\Omega = 0$.
Similarly, an \emph{infinitesimal symmetry}  of the structure $(E,\Lambda)$ is
a vector field $X$ on $\f M$ such that
$L_X E = [X,E] = 0$, $L_X\Lambda = [X,\Lambda] =  0$.

\bLm\label{Lm: i.1}
Let $X$ be a vector field on $\f M$. The following conditions
are equivalent:

1. $L_X\omega = 0$ and $L_X\Ome = 0$.

2. $L_X E= [X,E]=0$ and
$L_X\Lambda= [X,\Lambda]=0$.
\eLm

\bpf
For the dual structures we have the identities \eqref{Eq: 2.1}.
Then we get the identities
\begin{gather*}
i_{L_X\Lambda} \Omega + i_{\Lambda} L_X \Omega
 =
0\,,\quad
i_{L_XE} \omega + i_{E} L_X \omega
 =
0\,,
\\
i_{L_XE} \Omega + i_{E} L_X \Omega
 =
0\,, \quad
i_{L_X\omega} \Lambda + i_{\omega} L_X \Lambda
 =
0\,.
\end{gather*}
Suppose that $L_X\omega = 0$ and $L_X\Ome = 0$. Then 
 $i_{L_XE} \Omega= 0$ implies that $E$-horizontal part of $L_XE$ vanishes, and
$i_{L_XE}\omega = 0$ implies that $\omega$-vertical  part of $L_XE$
vanishes. So, $L_XE = 0$.

We have, \cite{JanMod09}, $\Omega\Fla\com \Lam\Sha = \id_{T^*M} - E\ten\omega$
which implies 
$$
(L_X\Omega)\Fla\com \Lambda\Sha + \Omega\Fla\com(L_X\Lambda)\Sha
= - L_X E\ten \omega - E\ten L_X \omega\,.
$$
So, $\Omega\Fla\com(L_X\Lambda)\Sha = 0$ which implies that $E$-horizontal part
of $(L_X \Lambda)\Sha(\alpha)$ vanishes  for any 1-form $\alpha$.
Further we have $i_\omega i_\alpha L_X\Lambda = 0$. Really,
\begin{align*}
i_\omega i_\alpha L_X\Lambda 
& =
i_{\alpha \wedge \omega} [X,\Lambda] 
=
i_{[X,\Lambda]} (\alpha \wedge \omega)
\\
& =
i_X d i_\Lambda (\alpha \wedge \omega) - 
i_\Lambda d i_X (\alpha \wedge \omega)
- i_{X\wedge\Lambda} d(\alpha \wedge \omega)
\\
& = - i_\Lambda [
d(\omega(X)) \wedge \alpha + \ome(X) \, d\alpha - d(\alpha(X)) \wedge \omega - \alpha(X) \, d\omega 
\\
& \quad 
+ (i_X d\omega)\wedge\alpha +  \alpha(X) \, d\omega - \ome(X) \, d\alpha + \omega\wedge(i_X d\alpha)]
\\
& = 
- i_\Lambda [(di_X \omega + i_X d\omega)\wedge\alpha ] = 
- i_\Lambda (L_X \omega \wedge\alpha)
= 0\,.
\end{align*}
So, we get that $\omega$-vertical part of $(L_X \Lambda)\Sha(\alpha)$ vanishes for any $\alpha$ and $(L_X \Lambda)\Sha (\alpha) = 0$. Then $L_X\Lambda = 0$.

\smallskip
The opposite implication can be proved in the same manner.
\hfill\QED
\epf

In differential geometry of odd-dimensional manifolds two types of geometrical
structures are usually considered: a {\em cosymplectic structure}
\cite{Albert89,dLS93,deLTuy96,LiMa87} is
defined by a {\em cosymplectic pair}
$(\ome,\Ome) \,,$
where 
$\ome$
is a 1--form and
$\Ome$
a 2--form, such that
$
d\ome = 0 \,,
$
$
d\Ome = 0 \,,
$
$
\ome \wed \Ome^n \nid 0 \,,
$
and a {\em contact structure} 
\cite[pag. 285]{LiMa87} is defined by a {\em contact pair}
$(\ome, \Ome) \,,$
where 
$\ome$
is a 1--form and
$\Ome$
a 2--form, such that
$
\Ome = d\ome \,,
$
$
\ome \wed \Ome^n \nid 0 \,.
$

More generally, we define an 
 {\em almost--cosymplectic--contact structure} 
 \cite{JanMod09} by a {\em regular almost--cosymplectic--contact
pair} 
$(\ome,\Ome) \,,$
where 
$\ome$
is a 1--form and
$\Ome$
a 2--form, such that
\bEq\label{Eq:a-c-c pair}
d\Ome = 0 \,,
\qquad
\ome \wed \Ome^n \nid 0 \,.
\eEq
Thus, an almost--cosymplectic--contact structure becomes a
cosymplectic structure when 
$d\ome = 0$ 
and a contact structure when
$\Ome = d\ome \,.$

We recall that a {\em Jacobi structure} \cite[pag. 337]{LiMa87} is
defined by a {\em Jacobi pair}
$(E, \Lam) \,,$
where
$E$
is a vector field and
$\Lam$
a 2--vector, such that
$
[E, \Lam] = 0 \,,
$
$
[\Lam, \Lam] = - 2 E \wed \Lam \,,
$
where
$[\,,]$
denotes the Schouten bracket.
Further, a {\em coPoisson structure\/} \cite{JanMod09} is defined by
a pair
$(E, \Lam)$
such that
$
[E, \Lam] = 0 \,,
$
$
[\Lam, \Lam] = 0 \,.
$

More generally, we define an {\em almost--coPoisson--Jacobi
structure} \cite{JanMod09} by an {\em almost--coPoisson--Jacobi pair}
$(E, \Lam) \,,$ 
along with a {\em fundamental 1--form} 
$\ome$
such that
\bEq\label{Eq:a-cP-J pair}
[E,\Lam] = 
- E \wed \Lam\Sha(L_E \ome) \,,
\qquad
[\Lam,\Lam] = 
2 \, E \wed \big((\Lam\Sha \ten \Lam\Sha) (d\ome)\big)
\,.
\eEq

We have 

\bTh\label{Th: 1.1}
{\rm \cite{JanMod09}}
 Let 
$(\ome,\Ome)$ 
be a regular pair and 
$(E,\Lam)$ 
its dual regular pair.
 Then,

\smallskip

{\rm (1)}
$(\ome,\Ome)$ 
is an almost-cosymplectic-contact pair if and only if 
$(E,\Lam)$ 
is an almost-coPoisson-Jacobi pair along with the fundamental 1-form 
$\ome \,;$

\smallskip

{\rm (2)}
$(\ome,\Ome)$ 
is a cosymplectic pair 
if and only if 
$(E,\Lam)$ 
is a coPoisson pair;

\smallskip

{\rm (3)}
$(\ome,\Ome)$ 
is a contact pair if and only if 
$(E,\Lam)$ 
is a Jacobi pair.
\hfill\END
\eTh

Note that all the above mentioned covariant and their dual contravariant
structures define generalized contact structures on $T\f M\oplus T^{*} \f M$ in
the sense of \cite{PooWad11,Wad12}.

\medskip
\textbf{Jets of submanifolds.}
Here we will recall the basics about jets of $1$-dimensional
submanifolds.  Our main sources are
\cite{ALV91,ManVit08,ModVin94}, where the structures constructed below are
given in the general case of jets of submanifolds of arbitrary
dimension.

Let $\f E$ be a manifold, and let $\dim \f E= 1+m$.  Let $r\geq 0$. An $r$-jet
of a $1$-dimensional submanifold $s:\f S \subset \f E$ at $x \in\f E$ is
defined to be the equivalence class of $1$-dimensional submanifolds having a
contact with $s$ of order $r$ at $x$. The equivalence class is denoted by
$j_rs(x)$, and the set of all $r$-jets of all $1$-dimensional submanifolds at
$x\in \f E$ is denoted by $J_{r\,x}(\f E,1)$. The set $J_{r}(\f
E,1)=\sqcup_{x\in\f E}J_{r\,x}(\f E,1)$ is said to be the $r$-jet space of
$1$-dimensional submanifolds of $\f E$. Of course we have $J_0(\f E,1)=\f
E$. For $r>p$ we have the projections $\pi^r_p:J_r(\f E,1) \to J_p(\f E,1)$.

A chart $(x^\lambda)$ on $\f E$ is said to be \emph{divided} if its coordinates
are divided in two subsets of $1$ and $m$ elements. In what follows Greek
indices $\lambda$, $\mu$, \dots will run from $0$ to $m$, and will label
coordinates on $\f E$, while Latin indices $i$, $j$, \dots will run from $1$ to
$m$. We will use the notation $(x^0,x^i)$ for divided charts.  We denote
by $(\der_\lambda)$ and $(d^\lambda)$ the local basis of vector fields and
$1$-forms induced by a chart $(x^\lambda)$.

A divided chart $(x^0,x^i)$ on $\f E$ is said to be \emph{adapted} to a
$1$-dimensional submanifold $s:\f S\subset \f E$ if $s$ can be expressed in
coordinates as $(\br x^0,s^i)$, where $\br x^0 \byd x^0|_{\f S}$ and $s^i\byd
x^i\com s \com (\br x^0)^{-1}$ are local real functions.  The set $J_r(\f E,1)$
has a natural manifold structure; a divided chart $(x^0,x^i)$ on $E$ induce the
chart $(x^0,x^i,x^i_{\ual})$ on $J_r(E,1)$ such that $x^i_{\ual}\circ s
=\der^{|\ual|}s^i/\der x^{\ual}$ (here $\ual$ is a multi-index of length $|\ual
|=r$ containing, in the $1$-dimensional case, only the index $0$ repeated $r$
times). The above charts yield a structure of a smooth manifold to $J_k(\f E,
1)$.  With respect to this smooth structure, the projections $\pi^r_p$ turn out
to be smooth bundles, and in particular $\pi^{r+1}_r$ turn out to be affine
bundles for $r\geq 1$. If $r=1$ then $J_1(\f E,1)=\operatorname{Proj(T\f E)}$,
the bundle of projective spaces obtained from each fibre of $T\f E \to \f E$.

Let $\f F$ be another manifold such that $\dim(\f E) \le \dim(\f F)$.
Let $f:\f E \to \f F$ be an immersion. Then $f$ maps (locally)
$1$-dimensional submanifolds of $\f E$ to $1$-dimensional submanifolds
of $\f F$ and preserves the contact, so we have the induced fibered
morphisms $J_rf:J_r(\f E,1) \to J_r(\f F,1)$, over $f$.
If we assume a vector field $X$ on $\f E$, then its flow $Fl^X_t$
is prolonged into the 1-parameter family of diffeomorphisms
$J_r(Fl^X_t):J_r(\f E,1) \to J_r(\f F,1)$ and we obtain the vector
field $X_{(r)} \byd \tfr {d}{dt}|_0 J_r(Fl^X_t)$,
called the \emph{$r$-jet lift} or the \emph{holonomic lift} of vector
fields on $\f E$ to vector fields on $J_r(\f E,1)$. Let us denote by
$\der_i^0$, $\der_i^{00}$, \dots the coordinate vector fields with respect to
the jet coordinates $x^i_0$, $x^i_{00}$, \dots. We have the
coordinate expression
\begin{align*}
X_{(1)} & = X^\lam \, \der_\lam + (\der_0 X^i
+ x^p_0 \, \der_p X^i - x^i_0 \, \der_0 X^0 - x^i_0 \, x^p_0
\, \der_p X^0) \, \der^0_i
\\
&= X^\lam \, \der_\lam + \br\del^i_\lam \,
\br\del^\mu_0 \, \der_\mu X^\lam \, \der^0_i\,.
\end{align*}
Here we use the notation $\br\del^i_\lam = \del^i_\lam - x^i_0 
\, \del^0_\lam$ and $\br\del^\mu_0 = \del^\mu_0 + x^p_0 
\, \del^\mu_p$. We have $[X_{(1)}, \ba X_{(1)}] = [X,\ba X]_{(1)}$.

We introduce the bundle $T_{1,0} \f E \byd J_{1}(\f E,1)\times_{J_r(\f E,1)}T\f
E$; the \emph{pseudo-hor\-i\-zon\-tal} subbundle $H_{1,0}\f E\subset T_{1,0}\f
E$ is defined by $H_{1,0}\f E\byd\{( j_{1}s(x) ,\upsilon) \in T_{1,0}\f E \mid$
$\upsilon \in T_{j_1s(x)} (j_1s(\f S))\}$.  We also have the
\emph{pseudo-vertical} bundle $V_{1,0}\f E\byd$ $T_{1,0}\f E/H_{1,0}\f E$.  The
bundles $H_{1,0}\f E$ and $V_{1,0}\f E$ are strictly related with the
horizontal and vertical bundle in the case of jets of fibrings.  A local basis
for the sections of the bundle $H_{1,0}\f E$ is
$D_0=\partial_0+x^i_{0}\partial_i$, while a local basis for the sections of its
dual $H_{1,0}{}^*\f E$ is $\odx{}^0\byd d^0|_{H_{1,0}\f E}$.  The fibred
inclusion $D\colon J_{1}(\f E,1)\to H_{1,0}{}^*\f E\otimes T_{1,0}\f E$ or,
equivalently, the fibred projection $\omega\colon J_{1}(\f E,1)\to
T_{1,0}{}^*\f E\otimes V_{1,0}\f E$ is said to be the \emph{contact structure}
on $J_{1}(\f E,1)$.  Their coordinate expressions are $D = \odx{}^0\otimes D_0
= \odx{}^0\otimes(\partial_0+x^i_{0}\partial_i)$ and $\omega=\omega^i\otimes
B_i$, where $\omega^i\byd d^i-x^i_{0}d^0$, and $B_i=[\der_i]$ is a local basis
of $V_{1,0}\f E$.  Note that $\omega^i$ annihilate $D_0$; they are said to be
\emph{contact forms}, and generate a space which is naturally isomorphic to
$V_{1,0}{}^*\f E$. Similar constructions can be repeated on $J_r(\f E,1)$ for
any order $r$ (see \cite{ManVit08} and references therein).

Lie derivatives can be used to characterize those vector fields on
$J_r(\f E,1)$ that are the holonomic lift of a vector field on $\f
E$. For instance, a vector field $Y\colon J_1(\f E,1)\to TJ_1(\f E,1)$
which projects onto a vector field $X$ on $T\f E$ fulfills $Y=X_{(1)}$
if and only if $L_{(X,Y)}D=0$ or, equivalently, $L_{(X,Y)}D_0=f^0D_0$,
where $f^0\colon J_1(\f E,1)\to\Rn$ is a smooth function.

A (second-order) \emph{differential equation} is a submanifold
$\mathcal{E}\subset J_2(\f E,1)$. A solution is a $1$-dimensional submanifold
$s$ of $\f E$ such that $j_2s\subset \mathcal{E}$. We will describe the
equation of particle motion in this way.

The calculus of variations can be formalized on jets of submanifolds
(\cite{Ded78,Vin78}; see also \cite{ManVit08}).  Given a $1$-form $\alpha$ on
$J_1(\f E,1)$ and a $1$-dimensional submanifold $s: \f L\subset \f E$ we have
the \emph{action} \cite{ManVit08} $A_{\f U}(s)=\int_{\f U} (j_1s)^*\alpha,$
where $\f U$ is a regular oriented open $1$-dimensional submanifold of $\f L$
with compact closure. 
Contact forms do not contribute to the action.
 The
\emph{horizontalization} \cite{ManVit08}
 takes the $1$-form $\alpha$ into a
section $h(\alpha)$ of $H_{1,0}^*\f E$ which does not contain contact factors,
hence $A_{\f U}(s)=\int_{\f U} (j_2s)^*h(\alpha)$.  If
$\alpha=\alpha_0d^0+\alpha_i d^i+\alpha_i^0d^i_0$ then we have the coordinate
expression $h(\alpha)=h(\alpha)_0\odx{}^0=
(\alpha_0+\alpha_ix^i_0+\alpha_i^0x^i_{00})\odx{}^0$.  There exists a natural
differential operator, the Euler--Lagrange operator $\C E$, bringing a
Lagrangian into its Euler--Lagrange expression.  We have $\C E(h(\alpha))\colon
J_3(\f E,1)\to V_{1,0}^*\f E \otimes H_{1,0}^*\f E$.  The corresponding
Euler--Lagrange equations are
\begin{equation}\label{Eq: 2.3}
  \C E(h(\alpha))=\left(\der_ih(\alpha)_0-D_0(\der_i^0
      h(\alpha)_0)+D_{00}(\der_i^{00}h(\alpha)_0)\right)
  \omega^i\otimes \odx{}^0=0,
\end{equation}
where $D_{00}=D_0^2$.
The Euler--Lagrange equation can be achieved in another way, using properties
of the $\mathcal{C}$-spectral sequence \cite{Ded78,Vin78}. Namely, it can be
proved that $\C E(h(\alpha))=h'(d\alpha)$, where $h'$ is horizontalization
followed by factorization modulo total divergencies.

\section{General relativistic mechanics}
\label{sec:gener-relat-phase}
\setcounter{equation}{0}

In this section we summarize the model of general relativistic phase
space by Jany\v ska and Modugno (see \cite{JanMod08} and references
therein).

\subsection{Phase space.} We assume the \emph{spacetime} to be a manifold $\f
E$, with $\dim \f E = 4$, endowed with a scaled Lorentz metric $g :\f E \to \B
L^2\otimes T^*\f E\otimes_{\f E} T^*\f E$ whose signature is $(-+++)$.
Moreover, we assume $\f E$ to be oriented and time oriented.  With reference to
a mass $m \in \B M \,,$ it is convenient to introduce the {\em rescaled
  metric\/} $G \byd \fr m\h g : \f E \to \B T \ten (T^*\f E \otimes_{\f E}
T^*\f E)$.  The associated contravariant tensors are $\ba g : \f E \to \B
L^{-2} \ten (T\f E \otimes_{\f E} T\f E)$ and $\ba G = \tfr \h m \ba g : \f E
\to \B T^* \ten (T\f E \otimes_{\f E} T\f E)$. In what follows, we will not
indicate obvious indexes in fibred (tensor) products.

A \emph{spacetime chart} is defined to be a divided chart
$(x^0,x^i)$
of
$\f E$
which fits the orientation of spacetime and such that the vector field
$\der_0$
is timelike and time oriented and the vector fields
$\der_i$
are spacelike. We have the coordinate expressions
$g = g_{\varphi\psi}d^\varphi\otimes d^\psi$,
$\ba g = g^{\varphi\psi} \der_\varphi \otimes \der_\psi$
where
$g_{\varphi\psi} \colon E \to {\B L}^2\otimes\Rn$
and
$g^{\varphi\psi} \colon E \to {\B L}^{-2}\otimes\Rn$
are mutually inverse matrices. Similarly
$G = G^0_{\lam\mu} \, u_0 \ten d^\lam \ten d^\mu$
and 
$\overline{G} = G_0^{\lam\mu} \, u^0 \ten \der_\lam \ten \der_\mu$
with
$G^0_{\lam\mu}, G_0^{\lam\mu}\colon \f E \to \Rn)$.

A time-like $1$-dimensional submanifold
$s:\f T\subset \f E$
is said to be a \emph{motion}, whose \emph{velocity} is
$j_1s$.

 For every arbitrary choice of a \emph{proper time origin}
$t_0 \in \f T$,
we obtain the \emph{proper time scaled function} given by
$\sig : \f T \to \baB T :
t \mto \fr1c \int_{[t_0, t]} \|\fr{ds}{d\br x^0}\| \, d\br x^0$.
 This map yields, at least locally, a bijection
$\f T \to \baB T$,
hence a (local) affine structure of
$\f T$
associated with the vector space
$\baB T$.
 Indeed, this (local) affine structure does not depend on the choice
of the proper time origin and of the spacetime chart.

 Let us choose a time origin
$t_0 \in \f T$
and consider the associated proper time scaled function
$\sig : \f T \to \baB T$
and the induced linear isomorphism
$T\f T \to \f T \car \baB T$.
 Moreover, let us consider a spacetime chart
$(x^0,x^i)$
and the induced chart
$(\br x^0)$
on
$\f T$.
Let us set
$\der_0 s^i \byd \fr{d s^i}{d\br x^0}$.
 The \emph{1st differential} of the motion
$s$
is the map
$ds \byd \fr{ds}{d\sig} : \f T \to \B T^* \ten T\f E$.
 We have
$g(ds, \, ds) = - c^2$
and the coordinate expression
\bEq\label{Eq: 3.1}
g(ds, ds) = - c^2,
\ \text{where}\ ds =
\fr{c_0 \,
u^0 \ten \big((\der_0 \comm s) + \der_0 s^i \, (\der_i \comm s)\big)}
{\sqrt{ |(g_{00} \comm s) +
2 \, (g_{0j} \comm s) \, \der_0 s^j +
(g_{ij} \comm s) \, \der_0 s^i \, \der_0 s^j|}} \,.
\eEq

The open subbundle $\M J_1 \f E\subset J_1(\f E,1)$ of velocities of
motions is said to be the (general relativistic) \emph{phase
  space}. In an analogous way we introduce the open subbundles $\M
J_r\f E \subset J_r(\f E,1)$.  Any spacetime chart $(x^0,x^i)$ is
adapted to each motion, hence the fibred chart $(x^0,x^i;x^i_0)$ on
$\M J_1\f E$ is global on the fibres of $\M J_1\f E \to \f E$.

The restriction of the pseudo-horizontal bundle $H_{1,0}\f E$ to $\M
J_1\f E$ will be denoted by $\M H_{1,0}\f E$; it admits the
trivialization $h_g\colon \M H_{1,0}\f E \to \M J_1\f E\times \baB T$, $(
j_1s(x) ,\upsilon)\to(j_1s(x),\pm\frac{\|{\upsilon}\|}{c})$,
where the sign depends on the time orientation of $\upsilon$.  Simple
computations show that the scaled vector field
\bEq\label{Eq: 3.3}
\K d \byd ((h_g{}^{-1})^*\otimes \id_{T\f E})\circ D|_{\M J_1\f E}
\colon \M J_1 \f E \to \B T^* \otimes T\f E
\eEq
has coordinate expression $\K d = c\,\alpha^0D_0$, where
$\alpha^0 \byd 1/\|D\|_0 = |g_{00} + 2g_{0j}x^j_0 + g_{ij}x^i_0x^j_0|^{-1/2}$.
The vector field $\K d$ is said to be the \emph{normalized contact
  structure}. It is characterised by the equality $\K d \circ j_1s =
ds$ for every motion $s$.  We have $g \circ (\K d,\K d) = -c^2$, hence
$\M J_1\f E$ can be regarded as a non-linear subbundle $\M J_1\f E
\subset \B T^*\otimes T\f E$ whose fibres are diffeomorphic to
$\Rn^3$.

\bRm
The property~\eqref{Eq: 3.1} says that $\M J_1\f E$ is a
\emph{universal mass shell}, \emph{i.e.} it represents a mass shell
for each choice of particle mass $m\in \B M$. At the same time,~\eqref{Eq:
  3.1} says that $\M J_1\f E$ coincides with the pseudosphere bundle
in~\cite{Iwa76}. The advantage of the jet space picture is that we can use
natural jet coordinates instead of dealing with a constraint, and at the same
time we give a geometric interpretation of projective coordinates used by
other authors~\cite{NiTo83}
\hfill\END
\eRm

We also have the dual counterpart of $\K d$
\begin{equation}\label{Eq: 3.4}
  \tau \byd -c^{-2}\, g\Fla \circ \K d \colon
  \M J_1\f E \to \B T \otimes T^{*}\f E,
\end{equation}
with coordinate expression $\tau = \tau_\lambda \, d^\lambda =
-c^{-1}\alpha^0 (g_{0\lambda} + g_{i\lambda}x^i_0) d^\lambda$.

The metric $g$ yields an orthogonal splitting of the tangent space $T
\f E$ on each $x \in \f E$ on which a time-like direction has been
assigned.  Hence, we have the splitting \cite{JanMod95,JanMod08}
\begin{equation}
  \label{Eq: 3.5}
  \M T_{1,0}\f E = \M H_{1,0} \udrs{\M
  J_1\f E} \M V_g\f E
\end{equation}
where $\M T_{1,0}\f E=T_{1,0}\f E|_{\M J_1\f E}$.  Let us introduce
the notation $\M V_{1,0}\f E\byd V_{1,0}\f E|_{\M J_1\f E}$. The projection
on $\M V_g\f E$ is denoted by $\theta$; it is readily
proved that $\theta=\id_{\M T_{1,0}\f E}-\tau\otimes\K d$.   We will use
the local basis $(D_0,N_i)$ adapted to the splitting~\eqref{Eq: 3.5}, where
$N_i\byd\partial_i-c\,\alpha^0\tau_iD_0$ is the basis of $\M V_g\f E$. Note
that $\omega|_{\M T_{1,0}}$ restricts to $\M V_g\f E$ to an
isomorphism $\M V_g\f E \to \M V_{1,0}\f E$; it reads in
coordinates as $N_i \to B_i$. The dual splitting of \eqref{Eq: 3.5} has a
local basis dual to $(D_0,N_i)$, namely $(N^0,\omega^i)$, where 
$N^0=d^0+c\,\alpha^0\tau_i\,\omega^i$ and $\omega^i$ are contact forms
(see the Preliminaries).

In what follows it is convenient to set $\breve{g}_{0\lambda}\byd
g(D_0,\partial_\lambda) =\br\del^\rho_0\, g_{\rho\lambda}$, $\ha g_{00} =
g(D_0,D_0)= \br\del^\rho_0\, \br\del^\sig_0\, g_{\rho\sig} = - (\alp^0)^{-2}$,
$\br g^{i\lam} = \ba g(\ome^i,d^\lam) = \br\del^i_\sig\, g^{\sig\lam}$.
We will use a similar notation for the rescaled metric $G$.

According to the above splitting we have the decomposition $g \circ \pi^1_0 =
g_\parallel+g_\perp$ of the lift of $g$ on $\M T_{1,0}\f E$, where $g\prl$ is a
negative-definite metric on $\M H_{1,0}$ and $g\per$ is a
Riemannian metric on $\M V_g\f E$.  The coordinate expression of the above
metric and of their contravariant form $\ba g\prl$ and $\ba g\per$ with respect
to the local basis $(D_0,N_i)$ and its dual $(N^0,\omega^i)$ are
\begin{alignat*}{2}
& g\prl{}_{00} = (\alpha^0)^{-2},
&&\overline{g}\prl{}^{00} = (\alpha^0)^2,
\\
&g\per{}_{ij}=g_{ij}+c^2\tau_i\tau_j,
\qquad &&\overline{g}_\perp{}^{ij} = g^{ij}-g^{i0}x^j_0-g^{0j}x^i_0+g^{00}x^i_0x^j_0.
\end{alignat*}
Note that the trivialization $h_g$ together with the splitting~\eqref{Eq: 3.5}
yields the further splitting $\M T_{1,0}\f E = \baB T \udrs{\M J_1\f E} \M
V_g\f E$; here the expressions of $g\prl$ and $\overline{g}\prl$ are
$g\prl{}_{00} = c^2$ and $\overline{g}\prl{}^{00} = c^{-2}$.

The vertical derivative $V\K d$ induces the linear fibred
isomorphisms
\begin{equation}
  \label{Eq: 3.8}
  v_g : V\M J_1\f E \to \B T^{*} \otimes \M V_g\f E\,,\qquad 
  v_g^{-1} : \B T^{*} \otimes \M V_g\f E\to V\M J_1\f E 
\end{equation}
over $\M J_1\f E$, with coordinate expressions $v_g = c\,\alpha^0\,d^i_0
\otimes N_i$ and $v_g^{-1} = \frac 1{c\,\alpha^0}\,\ome^i \otimes \der^0_i$.

We define a \emph{spacetime connection} to be a torsion free linear connection
$K : T\f E \to T^*\f E \ten TT\f E$ of the bundle $T\f E \to \f E$. Its
coordinate expression is of the type
\begin{equation*}
K =
d^\lam \ten 
(\der_\lam + K\col\lam\nu\mu \, \dot x^\mu \, \dt\der_\nu) \,,
\ssep{with}
K\col\mu\nu\lam = K\col\lam\nu\mu : \f E \to \Rn \,.
\end{equation*}

\smallskip

 We define
 a \emph{phase connection} to be a connection of the
bundle
$\M J_1\f E \to \f E \,.$
A phase connection can be represented, equivalently, by
$\Gam : \M J_1\f E \to T^*\f E \ten T\M J_1\f E \,,$
which is projectable over $\id_{T\f E}$,
or by $\nu[\Gam]=\id_{T\M J_1\f E}-\Gam :
\M J_1\f E \to T^*\M J_1\f E \ten V\M J_1\f E$,
or
$v_g[\Gam] \byd v_g \com \nu[\Gam] : \M J_1\f E \to
T^*\M J_1\f E \ten (\B T^* \ten \M V_g\f E)$.
 Their coordinate expressions are
\bgt
\Gam =
d^\lam \ten(\der_\lam + \Gam\Ga\lam i \, \der_i^0) \,,
\qquad
\nu[\Gam] =
(d^i_0 - \Gam\Ga\lam i \, d^\lam) \ten \der^0_i \,,
\\
v_g[\Gam] =
c \, \alp^0 \, (d^i_0 - \Gam\Ga\lam i \, d^\lam) \ten 
	 N_i \,,
\ssep{with}
\Gam\Ga\lam i : \M J_1\f E \to \Rn \,.
\end{gather*}

A spacetime connection $K$ induces naturally a phase connection
$\Gamma=\chi(K)$ \cite{JanMod08}, which is expressed in coordinates by
$\Gamma_{\varphi}{^i_0} = \br\del^i_\sig\, K{_\varphi}{^\sig}_\rho
\,\br\del^\rho_0$.

We denote by $K\Elg$ the \emph{Levi Civita connection}, i.e. the torsion free
linear spacetime connection such that $\nab g=0$. The connections $K\Elg$
and $\Gamma\Elg\byd \chi(K\Elg)$ are said to be \emph{gravitational}.

\subsection{Gravitational and electromagnetic forms.}
 Of course, being
$\dim \M J_1\f E=7$,
there exist no symplectic forms on
$\M J_1\f E$.
Hence, we are going to define a contact structure (in the sense of
\cite{LiMa87}) on
$\M J_1\f E$.
The gravitational connection
$\Gamma\Elg$
and the rescaled metric
$G$
induce the 2-form on
$\M J_1\f E$
\begin{equation}\label{Eq: 3.10}
\Omega\Elg \byd G_\perp\con(v_g[\Gamma\Elg])
\wedge \theta : \M J_1\f E \to
 \bigwedge^{2}T^{*}\M J_1\f E.
\end{equation}
It can be proved \cite{JanMod08} that $\Omega\Elg=-\fr{m\, c^2}{\h}\, d\tau$,
hence $\Omega\Elg$ is an exact form. Moreover, the form
$\tau\wedge\Omega\Elg\wedge\Omega\Elg\wedge\Omega\Elg$ is a scaled volume form
on $\M J_1\f E$, hence $\Omega\Elg$ is non degenerate.  It turns out that the
pair $(- \widehat\tau \byd - \fr{mc^2}{\h}\tau,\Omega\Elg)$ define on $\M
J_1\f E$ a contact structure.  $\Ome\Elg$ is said to be the \emph{phase
  gravitational 2-form} of our model.  We have the coordinate expression
\[
\Omega\Elg =  c_0\alpha^0 G^0_\perp{}_{ij} (d^i_{0} -
\Gamma\Elg{_{\varphi}}{^i_{0}}\,d^\varphi) \wedge \omega^j.
\]

Now, we assume the \emph{electromagnetic field} to be a closed scaled
$2$-form on $\f E$
\begin{equation}\label{Eq: 3.11}
  F :\f E \to (\B L^{1/2}\otimes\B M^{1/2}) \otimes \bigwedge^2
  T^*\f E.
\end{equation}
We have the coordinate expression
$F=2F_{0j}d^0\wedge d^j+F_{ij}d^i\wedge d^j$.
We denote a local potential of
$F$ with
$A :\f E \to T^*\f E$,
according to
$2 dA = F$.
In what follows we shall use the unscaled electromagnetic 2-form
$\wha F\byd (q/\hbar)\,F$ and its potential 
$\wha A\byd (q/\hbar)\,A$\,.

Given a charge
$q$,
the rescaled electromagnetic field
$ (q/2\hbar)\,F$
can be incorporated into the geometrical structure of the phase space,
\emph{i.e.}\ the gravitational form.  Namely, we define the
\emph{joined (total) phase 2-form}
\begin{equation}\label{Eq: 3.12}
  \Omega \byd \Omega\Elg + \frac{q}{2\hbar} F :\M J_1\f E \to 
  \bigwedge^2
   T^*\M J_1\f E.
\end{equation}
Of course $d \Omega = 0$ but $\Omega$ is exact if and only if $F$ is
exact. A (locally defined) $1$-form $\Theta$ such that $\Omega=d\Theta$ is said
to be a potential of $\Omega$. Every potential of $\Omega$ can be written as
$\Theta=- \widehat \tau+ \wha A$ up to a closed form. A potential $\Theta$ is
locally defined because the potential $A$ is locally defined. We have
$\tau \wedge \Omega \wedge \Omega \wedge \Omega = \tau \wedge
\Omega\Elg \wedge \Omega\Elg \wedge\Omega\Elg$, so $\Omega$ is non
degenerate. Hence, the pair $(- \, \widehat\tau,\Omega)$ is an
\emph{almost-cosymplectic-contact} (in the
sense of \cite{JanMod09}) encoding the gravitational and
electromagnetic (classical) structure of the phase space.  Note that, even
locally, $(- \, \widehat\tau+ \wha A,\Omega)$ cannot be regarded as a contact
pair, because the potential of $\Omega$ could vanish at some point.

We recall that a unique connection $\Gamma$
on $\M J_1\f E\to \f E$ can be characterized through the total 2-form
$\Omega$ \cite{JanMod08} by the formula \eqref{Eq: 3.10}. Namely the {\em
  joined (total) phase connection} $\Gam = \Gam\Elg + \Gam\Ele$,
where
$$\Gam\Ele \byd
- \fr{1}{2} v^{-1}_g \com G\Sha^2
\com \big(\wha F + 2\, \tau \wed (\K d\con\wha F)\big) \,
$$
with the coordinate expression
$\Gam\Ele =
- (1/(2c_0\alp^0)) \br G^{i\mu}_0 \,
(\wha F_{\lam\mu} - (\alp^0)^2 \br g_{0\lam} \,
\wha F_{\rho\mu} \, \br\del^\rho_0) \, d^\lam \ten \der^0_i$.

\subsection{Equation of particle motion.}
We can define a second order contact structure $\K d_2 : \M J_2\f E \to \B T^*
\uten{\M J_2\f E} T \M J_1\f E$. Then, the map $\K d \con \Gamma$ takes its
values in the subbundle $\M J_2\f E$.

We define a \emph{second order  dynamical connection}
$\gamma$
\cite{JanMod08} on spacetime as the map
\begin{equation}\label{eq:7}
\gamma \byd \K d \con \Gamma : \M J_1\f E\to \M J_2\f E
\overset{\K{d}_2}{\hookrightarrow} \B T^*\otimes T\M J_1\f E.
\end{equation}
The above map
$\gamma$
plays here a role analogous to that of the geodesic spray in the
formulation of Galilean mechanics on tangent spaces.
We have the coordinate expressions
\[
\gamma =
c\,\alpha^0(\partial_0+x^i_0\partial_i+\gamma_0{}^i_0\partial^0_i),
\qquad
\gamma_0{}^i_0\byd x^i_{00}\circ\gamma=\Gamma_0{}^i_0+\Gamma_j{}^i_0x^j_0.
\]

For the gravitational phase connection we have the {\em gravitational second
  order connection} $\gam\Elg \byd \K{d}\con \Gam\Elg$.
As a scaled vector field on $\M J_1\f E$, $\gamma\Elg$ fulfills
$\gamma\Elg\con \Omega\Elg=0$, $\gamma\Elg\con\tau=1$,
hence $\wha\gam\Elg \byd \tfr{\h}{m\,c^2}\,\gam\Elg$ is the Reeb vector 
field associated with the contact pair
$(- \, \widehat\tau,\Omega\Elg)$ \cite{LiMa87}.
The dual Jacobi structure is given by the pair
$(- \, \widehat\gam\Elg , \Lam\Elg )$, where $\Lam\Elg$ is
the {\em gravitational phase 2-vector}
given by
\bEq
\Lam\Elg \byd \ba G \con (\Gam\Elg\wed v_g^{-1}): \M J_1\f E \to \bigwedge^2
T\M J_1\f E
\eEq
with coordinate expression
$\Lam\Elg = (1/(c_0\alp^0)) \br G^{j\lam}_0\,(\der_\lam +
\Gam_\lam{}^i_0\,\der^0_i)\wed\der^0_j$.

It is natural to ask about analogous properties of the total pair
$( - \, \wha \tau,\Omega)$. It can be easily proved that there exists a second
order connection $\gamma$ such that $\gamma\,\lrcorner\,\Omega=0$ and
$\gamma\,\lrcorner\,\tau=1$,
i.e. $\widehat\gam \byd \tfr{\h}{mc^2}\gam$ is the Reeb vector field
associated with the almost-cosymplectic-contact pair
$( -\widehat\tau,\Omega)$ \cite{JanMod09}. Such a second order
connection takes the form
$\gamma=\gamma\Elg+\gamma^e$, where
\begin{equation}\label{eq:10}
\gamma^e\colon \M J_1\f E\to \B T^* \otimes V\M J_1\f E.
\end{equation}
Note that the above sum is performed in
$\M J_2 \f E$,
as
$\B T^*\otimes V\M J_1\f E$
is the associated fibre bundle \cite{ManVit08} to the affine bundle
$\pi^2_1 : \M J_2 \f E \to \M J_1 \f E$.
We have the coordinate expression
\[
\gamma^e= - \br G_0^{i\lam}(\wha F_{0\lam} +
\wha F_{j\lam}x^j_0) \, u^0 \ten \partial^0_i.
\]
Of course, $\gamma^e$ is the \emph{Lorentz force} associated with
$F$.

The \emph{equation of particle motion} is the submanifold of $\M J_2 \f E$
defined as follows
\begin{equation}\label{eq:8}
  \nabla[\gamma] = j_2s - \gamma\circ j_1s=0.
\end{equation}
In other words, the equation of motion is the image of the section
$\gamma$.  It is equivalent to $j_2s-\gamma\Elg\circ
j_1s=\gamma^e\circ j_1s$.  We have the coordinate expression
\begin{equation*}
  x^i_{00}- \br \del^i_\tau\, K{_\rho}{^\tau}{_\sig}
  \,\br\delta^\rho_0\, \br\delta^\sigma_0
  =- \frac{q}{m}\overline{g}_\perp{}^{ik}\, \br\delta^\rho_0\,F_{\rho k}.
\end{equation*}

\bRm
  The above system of equations is different from the usual equations
  of general relativistic mechanics. The main difference is that the
  above equations are on \emph{unparametrized trajectories} rather
  than on \emph{parametrized trajectories}.  Indeed, it is well-known
  that general relativistic mechanics is invariant with respect to
  reparametrizations \cite{FGT99,HenTei92}. The above approach
  allows us to discard the extra degree of freedom constituted by the
  parameter of motion. Hence, the equations of motion are just $3$
  instead of the standard $4$, at the cost of having a polynomial expression of
  degree $3$ in velocities $x^i_0$ instead of $2$, like in the standard
  parametrized geodesic equation. Mathematically, this corresponds to the
  fact that it is possible to regard $J_1(\f E,1)$ as the quotient
  $\reg T^1_1\f E/G^1_1$, where $\reg T^1_1\f E$ is the space of
  regular $1$-velocities of curves (here it coincides with the
  tangent bundle to $\f E$ without the zero section) and $G^1_1$ is
  the $1$-jet of reparametrizations of curves.\END
\eRm

The dual almost-coPoisson-Jacobi structure associated with
the almost-cosymplectic-contact structure $( - \, \widehat\tau,\Omega)$
is the pair $( - \,\widehat\gam,\Lam)$, where
the {\em joined (total) phase 2--vector\/}
$\Lam \byd \ba G \con (\Gam \wed v^{-1}_g) \,,$
splits as
$\Lam = \Lam\Elg + \Lam\Ele \,,$
where
$$
\Lam\Ele = \fr{1}{2} (v_g^{-1} \wed v_g^{-1})
(G\Sha(\tht^* (\wha F))) \,,
$$
i.e., in coordinates,
$\Lam\Ele = (1/(2(c_0 \alp^0)^2))
\br G^{i\lam}_0 \, \br G^{j\mu}_0 \,
\wha F_{\lam\mu} \, \der^0_i \wed \der^0_j$.

Let us remark that if we consider the unscaled 1-form 
$\widehat \tau =  \tfr{m\,c^2}{\h}\,\tau$ and the unscaled vector field
$\widehat\gam =  \tfr{\h}{m\,c^2}\,\gam $ we can
split the tangent and the cotangent bundles of the
phase space as
\bAl\label{Eq: tangent splitting} 
T\M J_1\f E 
 & = 
 H_{\gam}\M J_1\f E\oplus V_{\tau} \M J_1\f E
= \lang \widehat\gam\rang \oplus \ker(\widehat\tau)\,,
\\[2mm]
\label{Eq: cotangent splitting}
T^*\M J_1\f E 
& = 
H^*_{\tau}\M J_1\f E \oplus V^*_{\gam} \M J_1\f E
= \lang \widehat\tau\rang \oplus \ker(\widehat\gam)
\,,
\end{align}
where $H_{\gam} \M J_1\f E  \byd \lang \widehat\gam \rang$,
$V^*_{\gam} \M J_1\f E  \byd \ker (\widehat\gam)$,
$H^*_{\tau} \M J_1\f E  \byd \lang \widehat\tau \rang$,
$V_{\tau} \M J_1\f E  \byd \ker (\widehat\tau)$.
We have the mutually inverse isomorphisms 
$\Lam\Sha:V^*_\gam\M J_1\f E \to V_\tau\M J_1\f E$
and $\Ome\Fla: V_\tau\M J_1\f E \to V^*_\gam\M J_1\f E$.

The equation of motion can be derived from the variational
viewpoint, see \cite{ManVit05}; here we recall the main steps.
Taking the horizontal part of $\Omega$ we have $h(\Omega)=\eta$, where
$\eta\colon \M J_2\f E \to {\M V_g^*\f E}\otimes_{\M J_2\f E} {\M H_{1,0}^*\f E}$
has the coordinate expression
\[
  \eta =
  \frac{m}{\hbar}\left(c\,\alpha^0g_\perp{}_{ij}
      (x^i_{00}-\gamma\Elg_0{}^i_0)
    - \frac{q}{m} (F_{0j} +F_{ij}x^i_0)
  \right) \omega^j\otimes \odx{}^0.
\]
Indeed, the form $\Omega$ can be locally split as $\Omega=\eta+C$, where $C$ is
a contact form.  We observe that $\eta$ is a Euler--Lagrange morphism: it is
divergence-free because it takes values in a space of divergence-free forms
\cite{ManVit08}.  Now, the obvious link between $\eta$ and the equation of
motion is $\eta= \overline{g}_\perp (\nabla[\gamma])$. Of course, $\eta$ comes
from a Lagrangian: the relation $\Omega=d\Theta=d(-\widehat\tau+\widehat A)$
implies that $\eta$ admits the (local) Lagrangian $\C L\colon \M J_1\f E\to \M
H_{1,0}^*\f E$, $\C L=h(\Theta)$, with coordinate expression
\[
  \C L = L_0\,\odx{}^0 = \left(-\frac{mc}{\hbar\alp^0} + (\wha A_0+x^i_0\wha
    A_i)\right)\odx{}^0.
\]
Note that the condition $-\frac{mc}{\hbar\alp^0} + (\wha A_0+x^i_0\wha A_i) \ne
0 $ is equivalent to the fact that $ (-\wha\tau+\wha A)\wed\Ome^3 $ is a volume
form, and in this case $(\Theta,\Omega)$ is a contact structure on the region
of the phase space where $A$ is defined and $\Theta\neq 0$.  Of course, $\C L$
is defined on the same domain as $A$, and it is global if and only if $F$ is
exact (including the distinguished case $F=0$). A simple computation proves
that $\Theta$ is the Poincar\'e--Cartan form associated with the Lagrangian of
the above theorem.

The Hessian of $\C L$ is a non-singular matrix.  In particular we have
$\frac{\partial^2 L}{\partial x^i_0\partial x^j_0} = -\frac{mc}{\hbar}\,
\alpha^0 g_\perp{}_{ij}$.  Note that in the standard relativistic mechanics the
arc-length Lagrangian $\tilde{\C L}=\sqrt{g_{AB}\dot y^A\dot y^B}$ on $T\f E$
has an Hessian matrix whose rank is $3$, hence it is singular.  Indeed, we can
regard $\K d_{1} : \M J_1\f E\to\B T^{*}\otimes T\f E$ as a universal model of
all primary constraints which are characterized by~\ref{Eq: 3.1}. Such primary
constraints are called \emph{mass shells}, and are uniquely determined by the
choice of a time scale $u_0\in\B T$.

The Hamiltonian formalism is observer-dependent, so we introduce observers in
our setting. A section $o\colon \f E \to \M J_1\f E$ is said to be an
\emph{observer}; $o$ can also be interpreted as a vector-valued form
$o=\odx^0\otimes(\der_0+o^i_0\der_i)$, where $o^i_0=x^i_0\circ o$. An integral
motion of an observer is a motion $s$ such that $j_1s=o\circ s$. There exists
spacetime coordinates for which $o^i_0=0$; such coordinates are said to be
\emph{adapted to $o$}. Any observer $o\colon \f E\to \M J_1\f E$ yields a
projection of $T^*\f E$-valued forms onto $\M H_{1,0}\f E$. The \emph{observed
  Hamiltonian} $\C H[o]$ is defined to be the projection of the
Poincar\'e--Cartan form $\Theta$ onto $\M H_{1,0}\f E$ through $o$:
\begin{equation}\label{eq:14}
  \C H[o] = - o \con \Theta = - \left(\frac{mc}{\hbar}\,\alpha^0 (g_{00} +
  g_{i0}x^i_0) + \wha  A_0\right) \odx^0.
\end{equation}
The observed Hamiltonian and the (local) Lagrangian are connected by
the Legendre transformation as follows. Consider the \emph{momentum}
$\C P \byd V_{\f E}\C L\colon \M J_1\f E$ $\to V^*_{\f E}\M J_1\f E
\otimes \M H_{1,0} \f E\simeq \M V^*_{1,0}\f E$.
Then, by means of the inclusion $\omega\colon \M V^*_{1,0}\f E\to \M
T^*_{1,0}\f E$ and the observer we can introduce the \emph{observed
  momentum} $\C P[o] \byd (\omega\circ o) \con \C P$,
with coordinate expression (with respect to coordinates adapted to
$o$) $\C P[o] = \der^0_i L_0 d^i$. Then a simple computation shows
that the following identity holds:
\begin{equation}
  \label{eq:17}
  \C H[o] = h(\C P[o]) - \C L,\quad
  h(\C P[o]) = \der^0_i L_0 x^i_0\odx^0.
\end{equation}

\subsection{Examples}

In this section we consider two simple general relativistic spacetimes,
Minkow\-ski and Reissner--Nordstrom, and compute coordinate expressions for the
geometric objects of our model.

\paragraph{Minkowski spacetime.} We assume $\f E=\Rn^4$ and $g=l^2_0\otimes
g^0$, where $l^2_0$ is a unit of measurement of area and $g^0$ is the usual
Minkowski metric; we also assume $F=0$. We have the Cartesian coordinate
expressions:
\begin{align}\label{eq:2}
  &\alpha^0 = \left(\sqrt{|-1+(x^1_0)^2+(x^2_0)^2+(x^3_0)^2|}\right)^{-1},
  \\
  &\tau = -c^{-1}\alpha^0(-d^0+x^1_0d^1+x^2_0d^2+x^3_0d^3).
\end{align}
Note that all Christoffel symbols $K$ and $\Gamma$ are zero. Being
$G\per_{ij}=\frac{m}{\hbar}(\delta_{ij}+(\alpha^0)^2x^i_0x^j_0)$, it follows that
\begin{equation}\label{eq:3}
  \Omega=c_0\alpha^0\frac{m}{\hbar}\sum_{i,j=1}^3
  (\delta_{ij}+(\alpha^0)^2x^i_0x^j_0)d^i_0\wedge\omega^j,
\end{equation}
Note that in this case $\Theta=-\tau$ and $\gamma = \K d$.  Of course,
Euler--Lagrange equations reduce to $x^i_{00}=0$ even if the Lagrangian is
$L_0=-(mc)/(\hbar\alpha^0)$.  Since $M$ is parallelizable, its jet space is
also trivial: $J_1(\f E,1)=\f E \times \B P\Rn^4$ and $\M J_1\f E$ can be
identified with the subspace $\f E \times \Rn^3$ of time-like vectors. It
follows that we can choose the observer $o=0$ induced by the canonical basis of
$\Rn^4$. With respect to such an observer, the Hamiltonian takes the form $\C
H[o]=-(mc/\hbar)\alpha^0\odx^0$.

\paragraph{Reissner--Nordstrom spacetime.} We assume $\f
E=\Rn\times(\Rn^3\smallsetminus\{0\})$ with local coordinates
$(x^0,x^1,x^2,x^3)=(t,r,\theta,\varphi)$. We also assume a metric $g$ and an
electromagnetic field $\widehat A$, where $g=l^2_0\otimes g^0$, $l^2_0$ is a
unit of measurement of area and
\begin{align}\label{eq:12}
  g^0&=g_{00}dt\otimes dt+g_{11}dr\otimes dr+g_{22}d\theta\otimes d\theta +
  g_{33}d\varphi\otimes d\varphi
  \\
  &=-\left(1-\frac{k_s}{r}+\frac{k_q^2}{r^2}\right)dt\otimes dt+
  \left(1-\frac{k_s}{r}+\frac{k_q^2}{r^2}\right)^{-1}dr\otimes dr\notag
  \\
  &\hphantom{={}}
  +r^2(d\theta\otimes d\theta+\sin^2\theta d\varphi\otimes d\varphi).\notag
  \\
  \label{eq:13}
  \widehat A & =-(q_0/(\hbar_0 r))d^0.
\end{align}
Here coordinates on velocities are denoted by $r_0$, $\theta_0$, $\varphi_0$,
the constant $k_s$ is the Schwartzschild radius, $k_q$ is a characteristic
length of the Reissner--Nordstrom spacetime and $q_0$ and $\hbar_0$ mean the
numerical values of a charge $q$ and $\hbar$ with respect to the chosen units.
Then
\begin{align}\label{eq:4}
  & \alpha^0 = \left(\left|-\left(1-\frac{k_s}{r}+\frac{k_q^2}{r^2}\right)
    +\left(1-\frac{k_s}{r}+\frac{k_q^2}{r^2}\right)^{-1}r_0^2+
    r^2\theta_0^2+r^2\sin^2\theta^2\varphi_0^2\right|\right)^{-1/2},
  \\
  \label{eq:6}
  & \tau = -c^{-1}\alpha^0
  \bigg(-\left(1-\frac{k_s}{r}+\frac{k_q^2}{r^2}\right)dt
  \\
  & \hphantom{\tau = -c^{-1}\alpha^0\bigg(}
    +\left(1-\frac{k_s}{r}+\frac{k_q^2}{r^2}\right)^{-1}r_0dr
    +r^2\theta_0d\theta
    +r^2\sin^2\theta\varphi_0d\varphi\bigg),\notag
\end{align}
and  $G\per_{ij}=\frac{m}{\hbar}(g_{ij}+(\alpha^0)^2g_{ii}g_{jj}x^i_0x^j_0)$
(no sum in $i$, $j$), where we set $g_{ij}=0$ if $i\neq j$. In this case the
metric Christoffel symbols $K{}_\rho{}^\nu{}_{\sigma}$ are not all zero, their
expression can be easily computed.
It follows that
\begin{multline}\label{eq:5}
  \Omega\Elg=c_0\alpha^0\frac{m}{\hbar}\sum_{i,j=1}^3
  (g_{ij}+(\alpha^0)^2g_{ii}g_{jj}x^i_0x^j_0)
  \\
  (d^i_0-(K{}_\lambda{}^i{}_0-x^i_0K{}_\lambda{}^0{}_0
  +x^p_0K{}_\lambda{}^i{}_p-x^p_0x^i_0K{}_\lambda{}^0{}_p)d^\lambda)
  \wedge\omega^j.
\end{multline}
Note that $\widehat F=2d\widehat A=-(q_0/(\hbar_0 r^2))d^0\wedge d^r$.  The
Lagrangian is $L_0=-(mc)/(\hbar\alpha^0)-(q_0/\hbar_0 r)$. As before, $\f E$ is
parallelizable and there is a priviliged observer, with respect to which the
Hamiltonian takes the form $\C H[o]=-((mc/\hbar)\alpha^0g_{00} -(q_0/(\hbar_0
r)))\odx^0$.

\section{Infinitesimal symmetries of spacetime and phase objects}
\label{Infinitesimal symmetries of spacetime and phase objects}
\setcounter{equation}{0}

Symmetries of spacetime are diffeomorphisms which preserve the geometric
structure of $\f E$, \emph{i.e.}, the structure of pseudo-Riemannian manifold
and its consequences. On the other hand symmetries of the phase space are
fibered diffeomorphisms of the phase space which preserve geometrical objets on
the phase space, the time form and contact mappings.  These symmetries will be
the subject of the first subsection.  In the second subsection, we will
consider symmetries of connections and in the third subsection symmetries of
dynamical objects, \emph{i.e.}, the phase $2$-form and the phase $2$-vector.

\subsection{Spacetime and phase infinitesimal symmetries}
\label{Spacetime and phase infinitesimal symmetries}

We can define 3 types of symmetries of the spacetime $\f E$.

\bDf Let $f:\f E \to \f E$ be a diffeomorphism. Then $f$ is

1. a \emph{motion preserving  symmetry} of $\f E$ if $f$
transforms any motion in a motion, i.e. $g_{f(x)} (Tf(X),Tf(X)) < 0$
for any timelike vector $X\in T_x\f E$.

2. a \emph{time preserving symmetry} of $\f E$ if $f$ preserves the
time form, i.e. $\tau(j_1 s)(X) = \tau(j_1(f \com s)) (Tf(X))$.

3. a \emph{metric preserving  symmetry} of $\f E$, or an
\emph{isometry} of $\f E$, if $f$ is an isometry of the
pseudo-Riemannian manifold $(\f E, g)$.
\eDf

Any symmetry in the sense 1 preserves the phase space, in the sense that it can
be prolonged to a diffeomorphism $\M J_1f$ of $\M J_1\f E$.  Any symmetry in
the case 2 is a symmetry in the case 1 and any symmetry in the sense 3 is a
symmetry in the case 2 and 1.

A vector field on $\f E$ is said to be a \emph{motion} (respective \emph{time},
respective \emph{metric}) \emph{preserving infinitesimal symmetry} of $\f E$ if
its flow is a motion (respective time, respective metric) preserving symmetry
of $\f E$. Since $\M J_1\f E$ is defined by an `open' condition (note that
$\dim J_1(\f E,1)=\dim \M J_1 \f E$), every vector field on $J_1(\f E, 1)$
restricts to a vector field on $\M J_1\f E$ whose flow evidently consists of
motion preserving symmetries. This fact is stressed in the following Lemma.

\bLm
Any vector field $X$ on $\f E$ is a motion preserving infinitesimal
symmetry of $\f E$, i.e. its flow is motion preserving.
\eLm

\bpf
$\M J_1\f E$ is an open subspace in $J_1(\f E,1)$, so  
the holonomic lift $X_{(1)}$ of any vector field $X$ on $\f E$
can be restricted to the vector field on $\M J_1\f E$, 
denoted by the same symbol $X_{(1)}$. Then its flow $FL^{X_{(1)}}_t = \M J_1FL^X_t$ 
is a diffeomorphism of $\M J_1\f E$
which is the lift of a motion preserving diffeomorphism of $\f E$.
Hence the flow of $X$ is motion preserving. 
\QED
\epf

\bRm
The above Lemma shows a subtle difference between our approach and
the approach in \cite{Iwa76}: for the tangent lift of $X$ to $T\f E$ to restrict
to the pseudosphere bundle he needs $X$ to be an isometry.  Our lift is
different from the tangent lift (note also that $\dim T\f E=8$!), but we
recover isometries by requiring projectability and the symmetry of the
geometric structures of the phase space, see below.
\eRm

\bDf\label{df:phasesymm}
Let $Y$ be a vector field on $\M J_1\f E$ projectable on a vector field
$X$ on $\f E$. Then:

1. $X$ is a \emph{time preserving infinitesimal symmetry}  
of $\f E$ if its
flow preserves the time form, \emph{i.e.} $L_{X_{(1)}}\tau=0$.

2. $X$ is a \emph{metric preserving  infinitesimal symmetry} (or a {\em Killing
  vector field}) of $\f E$ if $X$ is an infinitesimal isometry of $g$,
\emph{i.e.} $L_Xg=0$.

3. $Y$ is a \emph{phase infinitesimal symmetry}
\eDf

Of course, case 2 implies case 1.

In what follows we shall characterize phase infinitesimal symmetries of the
basic phase objects, i.e. phase infinitesimal symmetries of $\K d$, $\tau$,
$\tht$, $v_g$ and $g\prl$. It is remarkable that even if we allow for
projectable infinitesimal symmetries we obtain that they must be the holonomic
lifts of distinguished infinitesimal symmetries of $\f E$.

\bRm
In principle we could study more general infinitesimal symmetries; we consider
this question in detail in Section~\ref{sec:perspectives}.
\eRm

\bLm\label{Lm: 0.1}
1. Let $\alpha$ be a $1$-form on $\f E$.
Then $\K{d} \con \alpha = 0$ if and only if
$\alpha = 0$.

2. Let  $\Phi$ be a spacetime 2--form. Then
$\K d\con \Phi =0$ if and only if $\Phi = 0$.

3. Let $\Psi$ be a $(0,2)$ symmetric tensor field on $\f E$.
Then $\Psi(\K{d},\K{d}) = 0$ if and only if
$\Psi = 0$.
\eLm

\bpf
Let $\alpha = \alpha_\lambda\, d^\lambda$.
Then 
$
\K{d} \con\alpha = c\,\alp^0\,
(\alpha_{0} + \alpha_{p}\, x^p_0 )\,
$
and
 $\K{d} \con\alpha = 0$ if and only if 
 $
  \alpha_{0} + \alp_{p}\, x^p_0 = 0\,
 $
 for all $x^p_0$, but, from $\alpha_\lambda :\f E\to\B R$, it is possible
 if and only if $\alpha_{\lam} = 0$. Hence 1 follows, and also 2 by a similar
 computation. If $\Psi=\Psi_{\lam\mu}\, d^\lam\ten d^\mu$, then 
$\Psi(\K{d},\K{d}) = (c\,\alp^0)^2\,
(\Psi_{00} + 2\, \Psi_{p0}\, x^p_0 + \Psi_{pq}\, x^p_0\, x^q_0)$, hence the
result follows.
\QED\epf

Now, we can characterize phase infinitesimal symmetries of basic phase objects.

\bTh\label{Th: IS}
Let $Y$ be a projectable vector field on $\M J_1\f E$ over
a vector field $X$ on $\f E$. Then, the following conditions are equivalent:

{\rm (1)} $X$ is a Killing vector field and $Y= X_{(1)}$.

{\rm (2)} $Y$ is a phase infinitesimal symmetry of $\K{d}$, i.e. $L_Y\K{d} =0$.

{\rm (3)}  $Y$ is a phase infinitesimal symmetry of $\tau$, i.e. $L_Y\tau =0$.

{\rm (4)} $Y$ is a phase infinitesimal symmetry of $\theta$, i.e. $L_Y\theta =0$.

{\rm (5)} $Y$ is a phase infinitesimal symmetry of $v_g^{-1}$, i.e. $L_Y v_g^{-1} =0$.

{\rm (6)} $Y$ is a phase infinitesimal symmetry of $g\prl$, i.e. $L_Yg\prl =0$.
\eTh

\bpf
(1) $\Leftrightarrow$ (2).
We define the Lie derivative of $\K d$ with respect to a projectable vector
field $Y$ on $\M J_1\f E$, over $X$ on $\f E$, as the generalized Lie
derivative by
\beq
L_Y \K d = T\K d \com Y - \mathcal{T}(X) \com \K d\,,
\eeq
where $\mathcal{T}(X) $ is considered as the vector field on
$\B T^* \ten T\f E$ obtained as the tensor product of the identity vector
field on $\B T^*$ and the tangent lift of $X$ to $T\f E$.
Then $L_Y \K d$ has the coordinate expression
\bal
L_Y \K d
& =
c \, \alp^0 \, \big[
\tfr 12 (\alp^0)^2 \, \brd\lam \, X^\rho \,\der_\rho\ha g_{00}
+ Y^p_0 \, \del^\lam_p
+ (\alp^0)^2 \, Y^p_0 \, \br g_{0p} \, \brd\lam
-  \br\del^\rho_0 \, \der_\rho X^\lam
\big] \, \der_\lam
\,
\end{align*}
and,
by taking into account the splitting~\eqref{Eq: 3.5}, we can rewrite $L_Y \K d$ as
\bAl\label{Eq: 4.1}
L_Y \K d
& =
c\,(\alp^0)^3 \, \big( 
\tfr 12 \, X^\rho \, \der_\rho \ha g_{00} 
+ \br g_{0\rho} \, \brd\sig \, \der_\sig X^\rho
\big) \, D_0
\\
& \quad\nonumber
+ c\, \alp^0 \, \big(
Y^i_0 - \br\del^i_\rho \, \brd\sig \, \der_\sig X^\rho
\big) \, N_i\,.
\end{align}
So $L_Y \K d = 0$ if and only if
\bAl\label{K d 5}
c\,(\alp^0)^3 \,\tfr 12 \,\brd\sig \,\brd\omega \, \big( 
 X^\rho \, \der_\rho g_{\sigma\omega} 
+  g_{\rho\omega} \,  \der_\sigma X^\rho
+  g_{\sigma\rho} \,  \der_\omega X^\rho
\big)
& = 
0\,,
\\[2mm] \label{K d 6}
 Y^i_0 - \br\del^i_\rho \, \brd\sig \, \der_\sig X^\rho
& =
0\,.
\end{align}
\eqref{K d 5} is equivalent with $(L_X g)(\K d, \K d) = 0$ and, by Lemma \ref{Lm: 0.1}, with $L_Xg = 0$.
Further, \eqref{K d 6} is equivalent with $Y = X_{(1)}$. 

\smallskip
(1) $\Leftrightarrow$ (3). From \eqref{Eq: 3.4} we have
\bEq\label{Eq: LYtau}
L_{Y} \tau = - \tfr 1{c^2} (L_{Y} \K d \con g + \K d \con
L_X g)\,.
\eEq

If $X$ is a Killing vector field and $Y = X_{(1)}$, then
$L_X g = 0$ and $L_Y\K{d} = 0$, i.e. $L_Y\tau =0$.

On the other hand let $L_Y \tau = 0$. 
Then 
$$
L_{Y} \K d \con g = - \K d \con
L_X g
$$
and, by using the isomorphism $g\Sha$, we get
$$
L_{Y} \K d = - (\K d \con L_X g)\Sha
= -c\, \alp^0\,\br\del^\rho_0\, (L_X g)_{\rho\sig}\, 
g^{\sig\lam}\,\der_\lam 
\,.
$$
In the adapted coordinates
\bEq\label{Eq: 4.7}
- (\K d \con L_X g)\Sha
= c\, (\alp^0)^3\,\br\del^\rho_0\,\br\del^\sig_0\, (L_X g)_{\rho\sig}\, D_0 - (c\, \alp^0)^2 \, \br\del^\rho_0\, (L_X g)_{\rho\sig}\, 
\br g^{\sig i}\, N_i\,. 
\eEq
Now, by comparing the $\K{d}$-horizontal parts of \eqref{Eq: 4.1}
and
\eqref{Eq: 4.7}, we get $L_X g =0$, i.e. $X$ is a Killing vector field.
By comparing the $\tau$-vertical parts of  \eqref{Eq: 4.1}
and
\eqref{Eq: 4.7} we get $Y=X_{(1)}$.

\smallskip
(1) $\Leftrightarrow$ (4).
We have $\theta=\id_{\M T_{1,0}\f E}-\tau\otimes\K d$\,, i.e.
\beq
L_Y\theta = - L_Y\tau\ten\K{d} - \tau\ten L_Y\K{d}\,.
\eeq

Let $X$ be a Killing vector field
and $Y = X_{(1)}$, 
 then $L_Y\tau =0$ and $L_Y\K{d} = 0$ which implies $L_Y\theta = 0$.

On the other hand let $L_Y\theta = 0$, then 
$\tau\ten L_Y\K{d} = - L_Y\tau\ten\K{d}$ which implies, by evaluating it on $\K{d}$, 
\beq
L_Y\K{d} = - (L_Y\tau)(\K{d})\,\K{d}\,.
\eeq
So, the $\tau$-vertical part of $L_Y\K{d}$ vanishes and by \eqref{Eq: 4.1} $Y = X_{(1)}$. Then the above equality reduces to
\bEq\label{Eq: LYd LYtau}
L_{X_{(1)}}\K{d} = - (L_{X_{(1)}}\tau)(\K{d})\,\K{d}\,,
\eEq
but, from \eqref{Eq: 4.1} and \eqref{Eq: LYtau}, we obtain
\bEq\label{Eq: Lhol}
L_{X_{(1)}}\K{d} = \tfr1{2\,c^2}\, (L_X g)(\K{d},\K{d})\, \K{d}\,,
\quad
L_{X_{(1)}}\tau = - \tfr1{c^2}\, \big( L_{X_{(1)}}\K{d}\con g + \K{d}\con L_X g\big)\,.
\eEq
\eqref{Eq: LYd LYtau} then leads to 
 $(L_Xg)(\K{d},\K{d})=0$, i.e., by Lemma \ref{Lm: 0.1}, $X$ is a Killing vector field. 

\smallskip
(1) $\Leftrightarrow$ (5).
We have the
coordinate expression
\bAl\label{Eq: 4.8}
L_Y v_g^{-1}
& =
- \tfr 1{c\,\alp^0} \big[
\tfr12 (\alp^0)^2\br\del^i_\mu \, X^\rho \, \der_\rho \ha g_{00} 
+ (\alp^0)^2\br\del^i_\mu \, Y^p_0 \, \br g_{0p} 
+ Y^i_0 \, \del^0_\mu
\\
& \quad \nonumber
+ \br\del^p_\mu \, \der^0_p Y^i_0 - \br\del^i_\rho \, \der_\mu X^\rho
\big] \, d^\mu \ten \der^0_i
\end{align}
and, in the adapted coordinates, we get
\begin{align*}
L_Y v_g^{-1}
& =
- \frac 1{c\,\alp^0} \big[
(Y^i_0 - \br\del^i_\rho\,\br\del^\mu_0\,\der_\mu X^\rho)\,N^0
- c\,\alp^0\, (Y^i_0 - \br\del^i_\rho\,\br\del^\mu_0\,\der_\mu X^\rho)\,\tau_j\,\ome^j
\\
&\quad
+
\big(\tfr12 (\alp^0)^2\del^i_j \, X^\rho \, \der_\rho \ha g_{00} 
+ (\alp^0)^2\del^i_j \, Y^p_0 \, \br g_{0p} 
+ \der^0_j Y^i_0 - \br\del^i_\rho \, \der_j X^\rho
\big)\, \ome^j \big] \, \ten \der^0_i\,.
\end{align*}
Then
$L_Y v_g^{-1} = 0$ if and only if 
\bAl
Y^i_0 - \br\del^i_\rho\,\br\del^\mu_0\,\der_\mu X^\rho 
& = \label{Eq: 4.10}
0\,,
\\
- c\,\alp^0\, (Y^i_0 - \br\del^i_\rho\,\br\del^\mu_0\,\der_\mu X^\rho)\,\tau_j
 + & \nonumber
\\
+
\big(\tfr12 (\alp^0)^2\del^i_j \, X^\rho \, \der_\rho \ha g_{00} 
+ (\alp^0)^2\del^i_j \, Y^p_0 \, \br g_{0p} 
+ \der^0_j Y^i_0 - \br\del^i_\rho \, \der_j X^\rho
\big)
& =\label{Eq: 4.11}
0\,.
\end{align}
Then \eqref{Eq: 4.10} is equivalent with $Y= X_{(1)}$ and \eqref{Eq: 4.11} reduces
to
\bAl
\tfr12 (\alp^0)^2\del^i_j \,\brd\rho\,\brd\tau\,( X^\sig \, \der_\sig g_{\rho\tau} 
+ g_{\tau\sig}\, \der_\rho X^\sigma + g_{\rho\sig}\, \der_\tau X^\sigma)
& =\label{Eq: 4.12}
0\,
\end{align}
which is equivalent with $X$ to be a Killing vector field.

\smallskip
(3) $\Leftrightarrow$ (6).
We have $g\prl = - c^2\, \tau\ten \tau$, then
$$
L_{Y}g\prl = - c^2 \big( L_{Y}\tau \ten \tau 
+ \tau \ten L_{Y}\tau \big):\M J_1\f E \to \B L^2\ten T^*\f E\ten T^*\f E
$$
and $L_{Y}\tau = 0$ implies  $L_{Y}g\prl = 0$.

 On the other hand $L_{Y}g\prl = 0$ implies  $L_{Y}\tau\ten\tau + \tau\ten
 L_Y\tau = 0$.
Now let us consider the splitting \eqref{Eq: 3.5} of $T\f E$
and the corresponding adapted base $(D_0,N_i)$.
We have $\tau(D_0) =  \tfr{1}{c\,\alp^0}$ and $\tau(N_i)=0$.
Then
\bal
0 &= (L_{Y}\tau\ten\tau + \tau\ten L_Y\tau)(D_0,D_0) =
\frac{2}{c\,\alp^0}\,(L_Y\tau)(D_0)\,,
\\
0 &= (L_{Y}\tau\ten\tau + \tau\ten L_Y\tau)(D_0,N_i) =
\frac{1}{c\,\alp^0}\,(L_Y\tau)(N_i)\,,
\end{align*}
i.e. $L_Y\tau = 0$ and $Y$ is an infinitesimal symmetry of $\tau$.
\QED\epf


\subsection{Infinitesimal symmetries of connections}
\label{Infinitesimal symmetries of connections}

Let us recall that linear spacetime connections $K$ on $\f E$ can be regarded
as sections of the natural bundle $\M K :Q\f E\to \f E$ of linear connections.
The bundle $Q\f E$ is defined as the affine subbundle of the bundle of linear
morphisms $K :T\f E \times_{\f E} T\f E\to TT\f E$ over $\id_{T\f E}$ such that
$(\tau_{T\f E},T\tau_{\f E})\circ K = \id_{T\f E\times_{\f E}T\f E}$. The Lie derivative of
$K$, with respect to a vector field $X$ on $\f E$, is then the vertical
vector field on $Q\f E$ defined as
\beq
L_X K = TK \com X - \mathcal{Q}(X)\,,
\eeq
where $\mathcal{Q}(X)$ is the flow lift, \cite[p. 59]{KMS93}, of $X$ on $Q\f E$.
Note that the vertical subspace of $TQ\f E$ is defined as $VQ\f E=\ker T\M
K$. In this way we obtain the standard Lie derivative with the coordinate
expression
\begin{multline*}
L_X K 
 = 
\big(
X^\rho \der_\rho K\col \mu\lam\nu -
	K_\mu{}^\rho{}_\nu \, \der_\rho X^\lam 
	+ K_\rho{}^\lam{}_\nu \, \der_\mu X^\rho
	+ K_\mu{}^\lam{}_\rho \, \der_\nu X^\rho
	- \der^2_{\mu\nu} X^\lam
	\big)
\\ d^\mu \ten \der_\lam \ten d^\nu \,,
\end{multline*}
where we used the natural identification $VQ\f E \simeq T^*\f E\ten T\f E\ten 
T^*\f E$.

Let us note  that if $X$ is an infinitesimal symmetry of $g$, then, by
naturality property, $X$ is the infinitesimal symmetry of $K\Elg$.

\bRm
Let us remark that we can define generally infinitesimal
symmetries of $K$ as projectable vector fields $Y$ on $T\f E$, over
$X$ on $\f E$, such that $L_Y K \byd TK\com Y - (\M T^*(X) \ten \M
T(Y)) \com K = 0$.  Then we have two distinguished cases.  First, if
$Y = \M T(X)$, then $L_{\M T(X)} K$ is identified with $L_X K$ in the above
sense. Second, if $Y = h^K(X)$, where $h^K(X)$ is the horizontal lift of $X$
with respect to the connection $K$, then $L_{h^K(X)} K = 0$ if and only if
$X\con R = 0$ where $ R $ is the curvature tensor of $ K $.
\END
\eRm

\bRm
Let us note that any spacetime  connection is of the form
$K = K\Elg + \Phi$ where $\Phi$ is a symmetric
(1,2)-tensor field on $\f E$.
Then a Killing vector field is an infinitesimal symmetry of $K$ if and only if
it is an infinitesimal symmetry of $\Phi$.
\hfill\END
\eRm

\smallskip

We define the Lie derivative of $\Gam$ with respect to a projectable vector
field $Y$ on $\M J_1\f E$, over $X$ on $\f E$, by
$$
L_Y \Gam = T\Gam \com Y - (\mathcal{T}^* (X) \ten \mathcal{T}(Y))
\com \Gam : \M J_1\f E \to
T^*\f E \ten V_{\f E}\M J_1\f E
\,,
$$ 
in coordinates, $L_Y \Gam = (X^\rho \, \der_\rho \Gam\Ga \lam i + Y^p_0 \,
\der^0_p \Gam\Ga \lam i + \Gam\Ga \rho i \, \der_\lam X^\rho - \der_\lam Y^i_0
- \Gam\Ga \lam p \, \der^0_p Y^i_0) \, d^\lam \ten \der^0_i$.  From $\nu[\Gam]
= \id_{T\M J_1\f E} - \Gam$ we have $L_Y \nu[\Gam] = - L_Y \Gam$, i.e.  $L_Y
\Gam = 0 $ if and only if $L_Y \nu[\Gam]= 0$.  Note that infinitesimal
symmetries of $\Gam$ and $v_g[\Gam]$ are not equivalent:
$L_Yv_g[\Gam] = L_Yv_g \circ \nu[\Gam] + v_g \circ L_Y\nu[\Gamma]$,
so $L_Y v_g =0$ and $L_Y\Gam =0$ imply $L_Yv_g[\Gam] = 0$.
Since $L_{X_{(1)}} \chi(K) = \br\del^i_\sig \, \br\del^\rho_0 \,
(L_X K)_\lam{}^\sig{}_\rho \, \, d^\lam \ten \der^0_i$, we obtain the following
result.

\bPr\label{Pr: infinitesimal symmetry K Gam[K]}
Let $K$ be a spacetime connection. Let 
$X$ be a vector field on $\f E$ and 
$X_{(1)}$ the corresponding holonomic lift. Then
\beq
L_X K = 0 \quad \Longrightarrow \quad L_{X_{(1)}}
\chi(K) = 0\,.
\eeq
In particular, any Killing vector field $X$ fulfills
$L_X K\Elg = 0$, $L_{X_{(1)}}\Gam\Elg = 0$, $L_{X_{(1)}}\gamma\Elg = 0$. 
\ePr

Let us remark that any phase connection is of the form $\Gam = \Gam\Elg +
\Sig$, where $\Sig:\M J_1\f E\to T^*\f E\ten V_{\f E}\M J_1\f E$ is a vertical
valued 1-form. Then it is easy to see that for a Killing vector field
$L_{X_{(1)}} \Gam = 0$ if and only if $L_{X_{(1)}} \Sig = 0$. Moreover, if $Y$
is an infinitesimal symmetry of $\K{d}$ then $L_Y\Gam =0$ implies
$L_Y\gam[\Gam]=0$. In fact, $L_Y \gam[\Gam] = 0$ if and only if
$L_Y\K{d}\con\Gam = - \K{d}\con L_Y\Gam$.

\bTh\label{Th: 4.25}
Let $X$ be a Killing vector field and $\Gam=\Gam\Elg + \Gam\Ele$ the total phase
connection. Then the following conditions are equivalent:
$$
 L_{X_{(1)}}\Gam =0 \quad \Leftrightarrow \quad L_{X_{(1)}}v_g[\Gam] = 0
 \quad \Leftrightarrow \quad L_{X}\wha F = 0 \quad \Leftrightarrow \quad
 L_{X_{(1)}}\gam[\Gam] = 0 \,. 
$$
\eTh

\bpf
For the
joined phase connection $\Gam = \Gam\Elg + \Gam\Ele$ we have
$L_{X_{(1)}} \Gam = 0$ if and only if   $L_{X_{(1)}} \Gam\Ele = 0$\,.
Similarly $L_{X_{(1)}} v_g[\Gam] = 0$ is equivalent with $L_{X_{(1)}} \nu[\Gam\Ele] = 0$, i.e.
$L_{X_{(1)}} \Gam\Ele = 0$. So we have to prove that 
$L_{X_{(1)}} \nu[\Gam\Ele] = 0$ is equivalent with $L_{X}\wha F = 0$.

We have $\Gam\Ele = - \tfr{1}{2} v^{-1}_g \com G\Sha^2
\com \big(\wha F + 2\, \tau \wed (\K d\con\wha F)\big)$ which implies that
 for a Killing vector field
$$
L_{X_{(1)}} \Gam\Ele 
 =
- \fr{1}{2}(v^{-1}_g \com G\Sha^2)\big( L_X\wha F +
 2\, \tau\wed(\K{d}\con L_X\wha F )
\big)\,.
$$
Then $L_X\wha F = 0$ implies $L_{X_{(1)}} \Gam\Ele = 0$ and $L_{X_{(1)}} v_g[\Gam] = 0$.

On the other hand
$L_X\wha F 
+ 2\, \tau\wed(\K{d}\con L_X\wha F) : \M J_1\f E\to T^*\f E \wed V^*_{1,0}\f E$
which implies that $(v^{-1}_g \com G\Sha^2)\big( L_X\wha F +
 2\, \tau\wed(\K{d}\con L_X\wha F ) = 0$ if and only if
$ L_X\wha F +
 2\, \tau\wed(\K{d}\con L_X\wha F) = 0$.
Evaluating values on the adapted base $(D_0,N_i)$
we get
\bal
0 
& =
(L_X\wha F + 2\, \tau\wed(\K{d}\con L_X\wha F))(D_0,D_0) =  (L_X\wha F)(D_0,D_0)\,,
\\
0 
& =
(L_X\wha F + 2\, \tau\wed(\K{d}\con L_X\wha F))(D_0,N_i) = 2 \,(L_X\wha F)(D_0,N_i)\,,
\\
0 
& =
(L_X\wha F + 2\, \tau\wed(\K{d}\con L_X\wha F))(N_i,N_j) = (L_X\wha F)(N_i,N_j)\,,
\end{align*}
hence $L_{X_{(1)}} \Gam\Ele = 0$ implies $L_X\wha F = 0$.

The implication $L_{X_{(1)}} \Gam = 0\Rightarrow L_{X_{(1)}}
\gam[\Gam]=0$ follows from the remarks preceding this Theorem.
On the other hand
\bal
0  & =  L_{X_{(1)}}
\gam[\Gam] =L_{X_{(1)}}
\gam[\Gam\Ele] =
\K{d}\con L_{X_{(1)}}
\Gam\Ele =
- \fr{1}{2} \K d\con (v^{-1}_g \com G\Sha^2)\big( L_X\wha F +
 2\, \tau\wed(\K{d}\con L_X\wha F )
\\
& = 
 - (v^{-1}_g \com G\Sha)\big(\K d\con L_X\wha F)\,.
\end{align*}
I.e. $L_{X_{(1)}}\gam[\Gam] =0$ is equivalent with  $\K{d}\con L_X\wha F =0$ and, by Lemma \ref{Lm: 0.1}, with $L_X\wha F =0$.
So $L_{X_{(1)}}\gam[\Gam] =0$ is equivalent with  $L_X\wha F =0$ which implies,
by Theorem \ref{Th: 4.25},
$L_{X_{(1)}} \Gam = 0$.
\QED\epf

\bRm
Let us note that $\gam\Elg$ is a scaled (non projectable) infinitesimal symmetry of
$\tau$. Really, we have $i_{\gam\Elg} \tau = 1$ and 
$d\tau = - \tfr \h{m\, c^2} \Ome\Elg$, Then $L_{\gam\Elg} \tau
= i_{\gam\Elg} d\tau + d i_{\gam\Elg} \tau = - \tfr \h{m\, c^2} i_{\gam\Elg} \Ome\Elg + d
(1) = 0$.
\END
\eRm

\subsection{Infinitesimal symmetries of dynamical objects}
\label{Infinitesimal symmetries of dynamical objects}

Now, let $\Ome = \Ome[G,\Gam]$ be the joined phase 2--form and $\Lam =
\Lam[G,\Gam]$ be the joined phase 2--vector given by the rescaled metric $G$
and the joined phase connection $\Gam$.  We define an infinitesimal symmetry of
$\Ome$ to be a vector field $Y$ on $\M J_1\f E$ such that $L_Y \Ome = 0$.  Let
us note that the joined $\gam$ is a (scaled non projectable) infinitesimal
symmetry of $\Ome$.  A phase function $f: \M J_1\f E\to\B R$ such that $\gam
. f = 0$ is said to be \emph{conserved}.

\bTh\label{Th: infinitesimal symmetry Ome conserved}
Infinitesimal symmetries  $Y:\M J_1\f E\to T\M J_1\f E$
of $\Ome$ are of local type $Y = \gam(\kap) + df\Sha$, where 
$\kap = \tau(Y):\M J_1\f E\to\ba{\B T}$ and $f:\M J_1\f E\to\B R$
is a conserved phase function determined up to a constant.
\eTh

\bpf
Let us consider a vector field $Y:\M J_1\f E\to T\M J_1\f E$.
Then, according to the tangent splitting 
$T\M J_1\f E = H_\gam \M J_1\f E\oplus V_\tau \M J_1\f E$,
we have $Y = \gam(\kap) + \ba Y$, where $\tau(\ba Y) = 0$. 

Then, by recalling the closure of $\Ome$ and $i_\gam\Ome =0 $, we
have 
$L_Y \Ome = di_Y\Ome = di_{\ba Y}\Ome$. I.e. $Y$ is an
infinitesimal symmetry of $\Ome$ if and only if $di_{\ba Y}\Ome = 0$
and hence, locally,
$i_{\ba Y}\Ome = df$ for a phase function $f$ and, moreover,
$0 = i_{\ba Y} i_\gam \Ome = i_\gam i_{\ba Y} \Ome =
i_\gam df =
\gam .f $.  Then
$\ba Y = df\Sha$ and $Y = \gam(\kap) + df\Sha$.
\QED
\epf

Infinitesimal symmetries of $\Lambda$ are vector fields $Y$ on $\M J_1\f E$
such that $ L_Y\Lambda = [Y,\Lambda] = 0 $. It is easy to realize that, in
general, infinitesimal symmetries of $\Omega$ and $\Lambda$ are not equivalent,
even in case of $Y$ projectable; they are equivalent only if we assume that
$Y=X_{(1)}$ for a Killing vector field $X$.

\bTh\label{Th: infinitesimal symmetry Gam Ome}
Let $\Gam$ be a phase connection,
$X$ be a Killing vector field and $X_{(1)}$ the
corresponding holonomic lift. Then
\begin{gather*}
 L_{X_{(1)}} \Gam = 0 \quad \Longrightarrow \quad
L_{X_{(1)}} \Ome = 0\,, \quad \quad L_{X_{(1)}} \Lam  = 0,
\\
L_{X_{(1)}}
\Ome = 0
\quad \Longleftrightarrow \quad
L_{X_{(1)}}
\Lam = 0.
\end{gather*}
\eTh

\bpf
For a Killing vector field $X$,  we get
$L_{X_{(1)}} \Ome = G\con ((v_g \com L_{X_{(1)}}\nu[\Gam]) \wed \tht)$, hence
$L_{X_{(1)}} \Gam =- L_{X_{(1)}} \nu[\Gam] =0$ implies $L_{X_{(1)}} \Ome = 0$,
and the same for $\Lam$.

It is easy to see that
$L_{X_{(1)}} \Lam = - (\Lam\Sha\ten\Lam\Sha)(L_{X_{(1)}} \Ome)$,
$L_{X_{(1)}} \Ome = - (\Ome\Fla\ten\Ome\Fla)(L_{X_{(1)}} \Lam)$, which yields
the result.
\QED
\epf

In view of Proposition~\ref{Pr: infinitesimal symmetry K Gam[K]}, if $K$ is a
spacetime connection and $X$ be a Killing vector field, then $L_X K = 0$
implies $L_{X_{(1)}} \chi(K) = 0$, hence $L_{X_{(1)}}\Ome =0$ and
$L_{X_{(1)}}\Lam =0$. Moreover, $L_{X_{(1)}} \Ome\Elg = 0$, $L_{X_{(1)}}
\Lam\Elg = 0$.


As a consequence of the above theorem, we can prove that symmetries of
dynamical structures are also symmetries of the Lagrangian and the
Euler--Lagrange morphism.

\bTh
Let $\Gam$ be a phase connection,
$X$ be a Killing vector field and $X_{(1)}$ the
corresponding holonomic lift. Then
\begin{enumerate}
  \item $L_{X_{(1)}} \Gam = 0 \quad \Longrightarrow \quad L_{X_{(1)}}\eta = 0
    \quad \Longleftrightarrow \quad L_{X_{(1)}}\C L = h(df)$;
  \item $L_{X_{(1)}}\C L = 0 \quad \Longleftrightarrow \quad L_{X_{(1)}}\Theta =
    0$;
  \item given an observer $o$ such that $L_{X_{(1)}}o=0$ then $L_{X_{(1)}}\C
    H[o]=0 \quad \Longleftrightarrow \quad L_{X_{(1)}}\C L = 0$.
\end{enumerate}
\eTh
\bpf
\begin{enumerate}
\item Recall that $\eta= \overline{g}_\perp (\nabla[\gamma])$. By $L_{X_{(1)}}
  \gamma[\Gamma]=0$ (Theorem~\ref{Th: 4.25}) and by
  $\overline{g}_\perp=\overline{g}(\theta,\theta)$, being $L_{X_{(1)}}\theta=0$
  we have $L_{X_{(1)}}\eta = 0$. The second part follows from the well-known
  fact \cite{Many99,Olv92} that the Euler--Lagrange operator $\C E$ commutes
  with the Lie derivative: $\C E(L_{X_{(1)}}\C L)=L_{X_{(1)}}\eta$. This
  implies that a symmetry of $\eta$ is a divergence symmetry of $\C L$,
  \emph{i.e.}, symmetry up to a total divergence, which is the
  horizontalization of a closed form.

\item Lie derivative with respect to a Killing vector field preserves the
  splitting~\eqref{Eq: 3.5}. Hence $L_{X_{(1)}} h(\Theta)=h(L_{X_{(1)}}
  \Theta)=(L_{X_{(1)}} \Theta)(\K d)$. Then, Lemma~\ref{Lm: 0.1} yields the
  result.

\item It is a straightforward consequence of the above facts.
\end{enumerate}
\epf

\section{Phase infinitesimal symmetries and special phase functions}
\label{Phase infinitesimal symmetries and special phase functions}
\setcounter{equation}{0}

From now we assume the almost-cosymplectic-contact structure and the dual
almost-coPoisson-Jacobi structure on the phase space given by the pairs $( - \,
\wha\tau,\Ome)$ and $( - \, \wha \gam,\Lam)$.  A \emph{phase infinitesimal
  symmetry of the classical phase structure} is defined to be a projectable
vector field $Y$ on $\M J_1\f E$ which is an infinitesimal symmetry of
$\wha\tau$ and $\Ome$.  We recall that, due to Lemma~\ref{Lm: i.1}, phase
infinitesimal symmetries of $( - \, \wha\tau,\Ome)$ and $( - \, \wha
\gam,\Lam)$ are the same.
 
\subsection{Special phase functions}
\label{Special phase functions}

In Theorem \ref{Th: infinitesimal symmetry Ome conserved} if we assume that $Y$
is a projectable vector field over a vector field $X$ on $\f E$ then in the
expression $Y = \gam(\kap) + df\Sha$ the phase function $f$ has to be a {\em
  special phase function}, and $Y$ is the so called {\em special Hamiltonian
  lift} $X\Upa[f]$ of $f$, \cite{JanMod06}.  $f$ is said to be a {\it generator
  of the infinitesimal symmetry $Y$} of $\Ome$.

We recall \cite{JanMod06} that special phase functions are of the type
\bEq\label{Special function}
f = - G(\K{d},X) + \br f = \wha\tau(X) + \br f\,,
\eEq
where $X$ is a vector field on $\f E$ and $\br f$ is a spacetime function.
So, the special phase function $f$ can be identified with the pair $(X,\br f)$.
We have the coordinate expression $f = - c_0 \alp^0 \br G^0_{0\rho} X^\rho +
\br f$.  We denote by $\spec(\M J_1\f E,\B R)$ the sheaf of special phase
functions.

The {\em special Hamiltonian lift} of a special phase function 
with respect to the joined almost-coPoisson-Jacobi pair $(-\widehat\gam, \Lam)$
is
\begin{equation}
X\Upa[f] = H[f] - \br f\, \widehat\gam
= df\Sha + (f-\br f)\, \widehat\gam \,
 =  df\Sha + \widehat\tau(X)\, \widehat\gam  \,,\label{eq:1} 
\end{equation}
where $H[f]= df\Sha + f\, \widehat\gam $ is the standard Hamiltonian lift. We
have the coordinate expression
\bAl
X\Upa [f]
& = 
X^{\lam} \, \der_\lam 
-  \br G^{i\sig}_0\, \big( - \tfr1{c_0\,\alp^0}\, \der_\sig\br f 
X^\rho\, \der_\rho \br G^0_{0\sig} + \br G^0_{0\rho}\,\der_\sig X^\rho
+ \tfr{1}{c_0\,\alp^0}\, X^\rho\,\wha F_{\rho\sig}\,\big) \, \der_i^0 \,.
\end{align}

For a phase function $f= - G(\K{d},X)$ corresponding to a vector field
$X$  on $\f E$ the special Hamiltonian lift coincides with the standard
Hamiltonian lift, i.e. $X\Upa[f] = H[f]$.
For a spacetime function $\br f$ the special Hamiltonian lift
is a vertical vector field.

Let us recall that we define the \emph{special bracket}
\cite{JanMod06} of two special phase functions 
$f = - G(\K{d}, X) + \br f $ and $h = - G(\K{d}, X') + \br h$ to be
\bEq
\db[f,h\db] = [f,h] - (\br f\,\widehat\gam.h 
- \br h\, \widehat\gam.f)\,,
\eEq
where $[,]$ is the Jacobi bracket.
We can rewrite the special brackets as 
\bal
	\db[f,h\db] 
& = 
	\{f,h\} + \big( (f- \br f)\,\widehat\gam.h 
- (h - \br h)\, \widehat\gam.f\big)
\\
& =
	\{f,h\} + \widehat\tau(X)\,\widehat\gam.h - \widehat\tau(X')\,\widehat\gam.f
\,,
\end{align*}
where $\{f,h\} = \Lam(df,dh)$ is the Poisson bracket.
The sheaf $\spec(\M J_1\f E, \B R)$ is closed with respect to the special
bracket \cite{JanMod06}. Indeed,
\bEq\label{Eq6.5}
\db[f,h\db]  = -G(\K{d},[X,X']) + X.\br h - X'.\br f + \widehat F(X,X')
\eEq
which is a special phase function given by the pair
$\big([X,X'], X.\br h - X'.\br f + \wha F(X,X')\big)$.

\subsection{Special phase functions and infinitesimal symmetries of
  dynamical objects}

The special Hamiltonian lift of a special phase function $f$
is an infinitesimal symmetry of $\Ome$ if and only if $f$
is conserved. Let us analyze more in detail this correspondence.

\bTh
The sheaf of conserved special phase functions is closed with respect to
the special bracket, and the special Hamiltonian lift is a Lie algebra
isomorphism from the Lie algebra of conserved special phase functions to the
Lie algebra of projectable infinitesimal symmetries of $\Omega$.
\eTh

\bpf
If $\wha\gamma.f = 0 $ and $\wha\gamma.h = 0 $ then
$\db[ f, h\db] = \{f,h\}$. Theorem then follows from
$$
\wha\gamma.\{f,h\} = \{\wha\gam.f,h\}
+ \{f, \wha\gam.h\} = 0\,.
$$
Indeed,
$\wha\gamma.\{f,h\} = (L_{\wha\gamma}\Lambda)(df,dh) +\Lam(L_{\wha\gamma}df, dh) + \Lam(df, L_{\wha\gamma}dh) =
- (\wha\gam\wed (L_{\wha\gamma}\wha\tau)\Sha)(df,dh)
+ \{\wha\gamma.f, h\} + \{f, \wha\gam.h\}$,
but $
(\wha\gam\wed (L_{\wha\gamma}\wha\tau)\Sha)(df,dh)
= i_{(L_{\wha\gamma}\wha\tau)\Sha}i_{\wha\gamma}(df\wed dh)
= (\wha\gamma.f) \, \Lambda(L_{\wha\gamma}\wha\tau,dh)
- (\wha\gamma.h) \, \Lambda(L_{\wha\gamma}\wha\tau,df)
= 0$.

We have, \cite{JanMod2012},
\bAl\label{Eq: 5}
X\Upa\big[\db[ f,h\db]\big]
 & = \big[X\Upa[f],X\Upa[h] \big]
 +
 \bigg(
 (\wha\gamma.f)\,\big(
 d\br h  - X' \con \wha F 
 \big)
 -  (\wha\gamma.h)\,\big(
 d\br f  - X \con \wha F 
 \big)
 \bigg)\Sha\,.
\end{align}
So for conserved special phase functions
$
X\Upa\big[\db[f,h\db]\big] = \big[X\Upa[f],X\Upa[h] \big]\,.
$
\QED\epf

\bPr\label{Pr: s.p.f. conserved}
A special phase function $f = - G(\K d, X) + \br f$ is
conserved if and only if
\beq
\K d . \br f - (X \con\wha F)(\K d) 
= \tfr 12 (L_X G)(\K d,\K d)\,.
\eeq
\ePr 

\bpf
We can calculate it in coordinates, by using the identity $\br G^{i\sig}_0 \, \br G^0_{i\rho}
= \del^\sig_\rho + (\alp^0)^2 \, \br g_{0\rho}\, \brd\sig$,
\bal
\gam . f 
& =
 i_\gam df 
=
c \, \alp^0 \, \brd\lam \, \big(
 \der_\lam \br f
 + \wha F_{\lam\sig} \, X^\sig
 - c_0 \, \alp^0 \, \tfr 12 \, \brd\mu \, 
  ( X^\rho \, \der_\rho G^0_{\lam\mu}
   + G^0_{\rho\mu} \, \der_\lam X^\rho
   + G^0_{\lam\rho} \, \der_\mu X^\rho
  )
\big)
\\
& =
\K d \con (d\br f - X \con \wha F - \tfr 12 \K d\con L_XG )
\,.
\QED
\end{align*}
\epf

The above proposition implies that, if $f = - G(\K d,X) + \br f \in \spec(\M
J_1\f E,\B R)$ where $X$ is a Killing vector field, then $f$ is conserved if
and only if $\K d . \br f - \, (X \con\wha F)(\K d)= 0$, i.e. if and only if $d
\br f = X \con\wha F$. This also implies that a spacetime function $\br f$ is
conserved if and only if it is constant.

\bLm\label{Lm: 5.6}
Let $X$ be a Killing vector field of $\f
E$ and
$f= - G(\K{d},X) +
\br f$ a corresponding conserved special phase function. Then 
\bEq
X\Upa[f] 
=
X_{(1)}\,.
\eEq
\eLm

\bpf
We have
\bal
X\Upa [f]
& = 
	X^{\lam} \, \der_\lam 
		-  \br G^{i\sig}_0\, \big(
- \tfr 1{c_0\, \alp^0}\der_\sig \br f
+ X^\rho\, \der_\rho \br G^0_{0\sig}
			+ \br G^0_{0\rho}\,\der_\sig X^\rho
			+ \tfr{1}{c_0\,\alp^0}\, 
			 X^\rho\,\wha F_{\rho\sig}\, 
		\big) \, \der_i^0
\\
& =
	X^{\lam} \, \der_\lam 
+ 		\br G^{i\sig}_0 \, \brd\tau \,
			\br G^0_{\sig\rho} \, \der_\tau X^\rho
	 \, \der_i^0   
 =
	X^{\lam} \, \der_\lam 
+ 		( \del^{i}_\rho - x^i_0\, \del^0_\rho)
			  \, \brd\tau \, \der_\tau X^\rho
	  \, \der_i^0
\\
& =
	X^{\lam} \, \der_\lam 
+ 		\br\del^{i}_\rho \, \brd\tau \, \der_\tau X^\rho
	  \, \der_i^0
	= X_{(1)}
\,. \QED
\end{align*}
\epf

Now, let us consider a special phase function $f$ satisfying $X\con \wha F
= d\br f$. Such a function generates an infinitesimal symmetry of
$\wha F$: $L_X\wha F=0$.

\bPr
Let $f$ and $h$ be special phase functions given by $(X,\br f)$, $(X',\br h)$
respectively, and let $X\con \wha F = d\br f$, $X'\con \wha F = d\br
h$. Then
\beq
  d(X.\br h - X'.\br f + \wha F(X,X')) = [X,X']\con\wha F \,,
\eeq
so that the special bracket of $f$ and $h$ generates an infinitesimal symmetry
of $\wha F$.
\ePr

\bpf 
We have
\bal
d(X.\br h - X'.\br f + \wha F(X,X'))
& = 
d(X\con X' \con \wha F - X'\con X \con \wha F + X'\con X \con \wha F)
\\
& = 
- d(\wha F(X,X'))\,.
\end{align*}
On the other hand
\bal
[X,X']\con\wha F 
& = \tfr 12 i_{[X,X']}\wha F
= \tfr 12 L_Xi_{X'}\wha F
= (di_X + i_Xd)(X'\con \wha F)
\\
& = 
d(X\con {X'}\con \wha F) + i_Xdd\br h
= - d(\wha F(X,X'))\,.\QED
\end{align*} 
\epf 
 
We shall call special phase functions satisfying the above properties {\em
  electromagnetic special phase functions}.

\bPr
If $f$ is a conserved electromagnetic special phase function, then its
corresponding vector field is a Killing vector.
\ePr

\bpf
If $\K d . \br f -  (X \con\wha F)(\K d)
= \tfr 12 (L_X G)(\K d,\K d)$ and $X\con \wha F = d\br f$ we get $(L_X G)(\K
d,\K d)=0$, i.e. $L_X G=0$, and conversely, due to Proposition~\ref{Pr:
  s.p.f. conserved}.
\hfill\QED\epf

\subsection{Holonomic special phase functions and infinitesimal
symmetries}
\label{Holonomic special phase functions and infinitesimal
symmetries}

We define the sheaf of {\em special holonomic phase functions} to be the
subsheaf of $f \in \spec(\M J_1\f E,\B R)$ such that $X\Upa[f] = X_{(1)}$.

\bTh
Let $X$ be a Killing vector field of $\f
E$ and
$f= - G(\K{d},X) +
\br f$ a corresponding special phase function. Then $f$ is conserved if and
only if it is holonomic.
\eTh

\bpf
The implication $\Rightarrow$ has been proved in Lemma \ref{Lm: 5.6}.

We have $X\Upa[f] = X_{(1)}$ if and only if
\beq
 -  \br G^{i\sig}_0\, \big( - \tfr1{c_0\,\alp^0}\,
\der_\sig\br f 
		+ X^\rho\, \der_\rho \br G^0_{0\sig}
			+ \br G^0_{0\rho}\,\der_\sig X^\rho
			+ \tfr{1}{c_0\,\alp^0}\, 
			 X^\rho\,\wha F_{\rho\sig}
\big) \, \der^0_i
= 
\br\del^i_\rho \, \brd\sig \, \der_\sig X^\rho\, \der^0_i\,.
\eeq
But we can write $\br\del^i_\rho = \br G^{i\sig}_0\, G^0_{\sig\rho}$
which imply that the above equation is equivalent to
\beq
- \tfr1{c_0\,\alp^0}\,\br G^{i\sig}_0\, \big( -  \der_\sig\br f 
	+  
			 X^\rho\,\wha F_{\rho\sig}
		+ c_0 \, \alp^0 \, \brd\tau\, (X^\rho\, \der_\rho G^0_{\tau\sig}
			+ G^0_{\tau\rho}\,\der_\sig X^\rho
+ G^0_{\sig\rho}\,\der_\tau X^\rho)
\big) \, \der^0_i
= 
0\,.
\eeq
Now, if we consider the splitting
$T^*\M J_1\f E = H^*_\tau\M J_1\f E \oplus V^*_{\gam}\M J_1\f E$, we get,
from the
identity 
$\br G^0_{i\rho} \, \br G^{i\sig}_0 = \del^\sig_\rho - c\, \alp^0 \, \brd\sig \,\tau_\rho$,
\beq
\der_\sig\br f 
-  X^\rho\,\wha F_{\rho\sig}
		= c_0 \, \alp^0 \, \brd\tau\, (X^\rho\, \der_\rho G^0_{\tau\sig}
			+ G^0_{\tau\rho}\,\der_\sig X^\rho
+ G^0_{\sig\rho}\,\der_\tau X^\rho) + k\, \tau_\sig\,,
\eeq
where $k \colon \M J_1\f E \to \B T^*$ given as
$k = \K d.\br f - (X\con\wha F)(\K d) - (L_XG)(\K d,\K d)$\,.

Now, let us suppose that $f$ be holonomic. By
Proposition~\ref{Pr: s.p.f. conserved} we get
$$
d\br f - X\con \wha F = (\K d.\br f - (X\con\wha F)(\K d))\,\tau
$$
which is possible if and only if $ d\br f - X\con \wha F = 0$\,,
i.e. if and only if $f$ is conserved.
\hfill\QED\epf

Of course, a metric special phase function $f$ is holonomic and conserved if
and only if $d\br f = X\con\wha F$. In Lemma \ref{Lm: 5.6} we have proved that
for a Killing vector field $X$ if a corresponding special phase function $f= -
G(\K{d},X) + \br f$ is conserved then it is holonomic. Now we shall prove the
equivalence.

\bTh\label{Th: IS F}
Let $f = - G(\K d,X) + \br f$ be a metric special phase
function. Then $f$ is conserved and holonomic if and only if $X$ is an
infinitesimal symmetry of $\wha F$.
\eTh

\bpf
A metric special phase function is conserved and holonomic if and only
if $d\br f = X \con\wha F$, i.e. $X\con\wha F$ is closed.
Then, for closed $\wha F$, $0 = d(X\con\wha F) = \tfr12 \, L_X\wha F$
and $X$ is an infinitesimal symmetry of $\wha F$.

On the other hand, if  $X$ is a Killing vector field which is an infinitesimal
symmetry of $\wha F$, then
$0 = L_X \wha F = 2 \, d(X\con\wha F)$ and (locally) $X\con\wha F = d\br f$, 
where $\br f$ is a spacetime function given up to a constant.
Then $f = - G(\K d,X) + \br f$ is a conserved and holonomic 
special phase function.
\hfill\QED
\epf

We call $f\in \spec (\M J_1\f E, \B R)$,  $f = - G(\K d, X) + \br f$,
\emph{self--holonomic} if $i_{X_{(1)}}\Ome = df$. 

\bTh\label{Th: 5.13}
$f$ is a self-holonomic special function if and only if $f$ is conserved and
holonomic.
\eTh

\bpf
If $f$ is a self--holonomic function, then $f$ is conserved. Indeed,
let $f$ be self--holonomic; then
$\gam. f = i_\gam df = i_\gam i_{X_{(1)}} \Ome =  - i_{X_{(1)}}i_\gam
\Ome = 0$. Moreover, $X\Upa[f] = \gam(\tau(X)) + df\Sha = 
\gam(\tau(X)) + \Lam\Sha(\Ome\Fla(X_{(1)}))
=\gam(\tau(X)) + X_{(1)} - \gam(\tau(X))=X_{(1)}$.

On the other hand let $f$ be a conserved special holonomic function. Then
\beq
i_{X_{(1)}}\Ome = i_{X\Upa[f]} \Ome = i_{(df\Sha + \gam(\tau(X))}\Ome
= (\Ome\Fla \com \Lam\Sha)(df) = df - \gam(df)\,\tau
= df - (\gam .f)\, \tau = df. \QED
\eeq
\epf

\bCr
Let $X$ be a Killing vector field and $f = - G(\K d,X) +\br f$ 
a corresponding
special phase function. Then the following conditions are equivalent:

1) $f$ is conserved; \qquad
2) $f$ is holonomic; \qquad
3) $f$ is self--holonomic.
\END
\eCr

\bTh\label{Th: self-holonomic IS Ome}
1)
Let $f = - G(\K d,X) + \br f $
be a self--holonomic special phase function. Then,  
$X_{(1)}$ is an infinitesimal symmetry of $\Ome$.

2)
Let $X$ be a vector field on $\f E$ such that 
$X_{(1)}$ is an infinitesimal symmetry of $\Ome$.
Then there exists a unique (up to a constant)  
self--holonomic special phase function $f$ such that
$X_{(1)} = X\Upa[f]$.
\eTh

\bpf
1) 
Let $f = - G(\K d, X) + \br f$ be a self--holonomic special phase function,
i.e. $i_{X_{(1)}} \Ome = df$. Then
$L_{X_{(1)}} \Ome = d i_{X_{(1)}} \Ome = ddf = 0$.

2) 
Let $X_{(1)}$ is an infinitesimal symmetry of
$\Ome$. Then by Theorem 
\ref{Th: infinitesimal symmetry Ome conserved} $X_{(1)}$ is 
of the form $X_{(1)} = \gam(\tau(X)) + df\Sha$,
where $f$ is a conserved phase function uniquely given up to a constant. 
Projectability on $X$ then
implies that $f$ has to be a special phase function 
$f = - G(\K d,X) + \br f$. Moreover, we have 
$
X_{(1)} = X\Upa[f] \,.
$
\QED
\epf

\bPr\label{Pr: 5.18}
The subsheaf of self-holonomic special phase functions
is closed with respect to the special bracket, i.e. if $f$, $h$ are
self-holonomic, with $f = -G(\K d,X) + \br f$, $h=-G(\K d,X') + \br h$, then
\beq
i_{[X,X']_{(1)}}\Ome = d\db[f,h\db]\,.
\eeq
\ePr

\bpf
We have $\gam.f = \gam .h =0$ and $X_{(1)} = X\Upa[f]$,
$X'_{(1)} = X\Upa[h]$.
To prove that $i_{[X,X']_{(1)}}\Ome = d\db[f,h\db]$ it is sufficient to prove that
$(\Lam\Sha\com \Ome\Fla)([X,X']_{(1)}) = \Lam\Sha(d\db[f,h\db])$ and
$i_\gam i_{[X,X']_{(1)}} \Ome = i_\gam d \db[f,h\db]$.
We have
\bal
(\Lam\Sha\com \Ome\Fla)([X,X']_{(1)}) 
& =
[X,Y]_{(1)} - \tau([X,X'])\gam 
= [X_{(1)}, X'_{(1)}] -  \tau([X,X'])\gam 
\\
& =
\big[X\Upa[f],X'{}\Upa[h]\big] - \tau([X,X'])\gam 
= 
X\Upa\big[ \db[ f, h\db]\big] - \tau([X,X'])\gam 
\\
& =
\Lam\Sha (d\db[ f,h \db] ) + \tau([X,X'])\gam - \tau([X,X'])\gam \,.
\end{align*}
Further we have $i_\gam i_{[X,X']_{(1)}} \Ome = - i_{[X,X']_{(1)}} i_\gam \Ome = 0$,
and 
\beq
i_\gam d \db[f,h\db] = i_\gam d \{f,h\} = \{\gam.f,h\} + \{f,\gam.h\} = 0\,.
\QED
\eeq
\epf

\subsection{Characterization of infinitesimal symmetries of the classical
phase structure}
\label{Infinitesimal symmetries of the classical phase structure}

We shall start with infinitesimal symmetries of the contact gravitational phase
structure given by the pair $(-\,\wha\tau,\Ome\Elg)$.

\bTh\label{Th: ISCGS} All projectable infinitesimal symmetries of the classical
contact gravitational structure are, equivalently:
\begin{enumerate}
\item holonomic lifts of Killing vector fields;
\item vector fields $Y=(d\widehat{\tau}(X))\Sha +
  \widehat\tau(X)\,\widehat\gamma\Elg$, where $X$ is a spacetime vector field
  such that $\widehat \gamma\Elg.(\widehat{\tau}(X)) = 0$.
\end{enumerate}
\eTh

\bpf
  Item 1 follows from Theorem~\ref{Th: IS} and
  Proposition~\ref{Pr: infinitesimal symmetry K Gam[K]}.
  Item 2 has been proved in \cite{Jan11}. However, this statement follows from
  the equivalence between item 1 and item 2. Indeed, it is enough to observe
  that $\widehat{\gamma}\Elg.\widehat{\tau}(X) = -
  (1/c^{2})(L_{X}g)(\K d,\K d)$. This implies the equivalence.
  \QED
\epf

Now, let us assume the total almost-cosymplectic-contact pair
$(-\,\wha\tau,\Ome=\Omega[G,F])$.

\bTh\label{Th: 5.17} All projectable infinitesimal symmetries of the classical
almost-co\-symplectic-contact structure $(-\,\wha\tau,\Ome)$ are, equivalently:
\begin{enumerate}
\item special Hamiltonian lifts of (conserved) special phase functions $f = -
  G(\K d,X) + \br f$, where $X$ is a Killing vector field of $\f E$ and $d\br f
  = X\con\wha F$;
\item vector fields $ Y=df\Sha + \widehat \tau(X)\,\widehat \gamma$ where $f$
  is a (conserved) special phase function $f= - G(\K{d},X) + \breve{f}$ such
  that $\widehat \gam.f = 0$ and $- i_{df\Sha}\Omega\Elg -
  \widehat{\tau}(X) \, i_{\widehat{\gamma}\Ele}\, \Omega\Elg +
  d(\widehat{\tau}(X)) = 0$ is satisfied.
\end{enumerate}
\eTh
\bpf
From Theorem
\ref{Th: IS} it follows that infinitesimal symmetries of $\wha\tau$ 
are  holonomic lifts $X_{(1)}$ of Killing vector fields of $\f E$. 
By Theorem \ref{Th: self-holonomic IS Ome} $X_{(1)}$ are infinitesimal
symmetries of $\Ome$ if and only if they are special Hamiltonian lifts of
self-holonomic (metric) special phase functions.
Item 1 now follows from Theorem~\ref{Th: IS F}. Item 2 has been proved in
\cite{Jan11}, but it also follows from the equivalence of the two statements.
 The equivalence follows from the fact that for conserved
special phase functions  the condition 2 can be reduced to
$ - d \breve{f} + X \con \wha F   = 0$. In fact,
$i_{\widehat{\gamma}\Ele}\, \Omega\Elg = - \wha{\K{d}} \con \wha F$
and
\bal
i_{df\Sha}\Omega\Elg
& =
(\Omega\Elg)\Fla\circ (\Lambda\Elg + \Lambda\Ele)\Sha(d(\wha\tau(X))+d\br f)
\\
& = 
d(\wha\tau(X))+d\br f - \wha\gamma\Elg(d(\wha\tau(X)) + d\br f)\,\wha\tau
+ (\Omega\Elg)\Fla\circ ( \Lambda\Ele)\Sha(d(\wha\tau(X))).
\end{align*}
But from $\wha\gamma.f =0$ we have
$\wha\gamma\Elg(d(\wha\tau(X)) + d\br f) = - \wha\gamma\Ele(d(\wha\tau(X)) +
d\br f) = - \wha\gamma\Ele(d(\wha\tau(X))) = \wha{\K{d}}\con( X \con \wha F)$.
Moreover, $(\Omega\Elg)\Fla\circ ( \Lambda\Ele)\Sha(d(\wha\tau(X)))
=
- X \con \wha F + (\wha{\K{d}}\con \wha F)\, \wha\tau(X)
+ \wha{\K{d}}\con (X\con \wha F )\,\wha\tau$,
from which the equivalence follows.\QED
\epf

\bCr
Infinitesimal symmetries of the classical structure are special Hamiltonian
lifts $X\Upa[f]$ of special phase functions $f= - G(\K d,X) + \br f$, where
$L_X G = 0$ and $d\br f = X\con\wha F$.
In this case $X\Upa[f] = X_{(1)}$ and the coordinate expression
of infinitesimal symmetries of the classical structure are
$X_{(1)} = X^\lam \, \der_\lam + \br\del^i_\rho \, \brd\sig \, \der_\sig X^\rho
\, \der^0_i$, where $\der_\lam\br f  = X^\rho\,\wha F_{\rho\lam}$ and
$X^\rho \, \der_\rho G^0_{\lam\mu}
+ G^0_{\rho\mu} \, \der_\lam X^\rho
+ G^0_{\lam\rho} \, \der_\mu X^\rho
 = 0$.\END
\eCr

\bTh
Special phase functions which are generators of phase
infinitesimal symmetries of the classical structure form a Lie algebra with
respect to the special bracket.
\eTh
\bpf
By Theorem \ref{Th: 5.17} generators of infinitesimal symmetries of the
classical structure are self-holonomic metric special phase functions. Theorem
now follows from Proposition \ref{Pr: 5.18}.
\hfill\QED
\epf

Note that for a Killing vector field $X$ the generator of the infinitesimal
symmetry $X\Upa[f]$ of the classical structure is a pair $(X,\br f)$, where
$\br f$ is a spacetime function satisfying $d\br f = X\con \wha F$, i.e. $\br
f$ is given up to a constant.

\subsection{Momentum map and special phase functions}
\label{Momentum map and special phase functions}

In this subsection, we will extend the theory of momentum map from symplectic
geometry (see \cite{OrRa04}) to our almost-cosymplectic-contact geometry. The
extension mimick previous more general theories for both contact and
cosymplectic geometry \cite{Albert89,CC02,dLS93}. We prove that our model
always admits a momentum map.

Let $G$ be a Lie group and $\Phi\colon G\times \M J_1\f E \to \M J_1\f E$ be a
left action such that for each $g\in G$ $\Phi_g^*\widehat\tau=\widehat\tau$ and
$\Phi_g^*\Ome=\Ome$. The associated infinitesimal action
$\phi\colon\mathfrak{g}\times \M J_1\f E \to T\M J_1\f E$, where
$\phi(\xi)_{j_1s(x)}= -T \Phi_{j_1s(x)}(\xi)$ is an action by infinitesimal
symmetries of the classical almost-cosymplectic-contact structure
$(-\,\wha\tau,\Ome)$. If we suppose $\Phi$ or $\phi$ to be projectable to
actions $\bar\Phi$ and $\bar\phi$ on $\f E$, then by Theorem~\ref{Th: 5.17} for
each $\xi\in\mathfrak{g}$ we have $\phi(\xi)=\bar\phi(\xi)_{(1)}$, where
$\bar\phi(\xi)$ is a Killing vector field which is also a symmetry of the
electromagnetic field $F$.

A \emph{momentum map} for a projectable action $\Phi$ of a Lie group $G$ of
infinitesimal symmetries of the classical almost-cosymplectic-contact structure
is a map $J\colon \M J_1\f E \to \mathfrak{g}^*$ such that $i_{\phi(\xi)}\Omega
= -dJ_\xi$, where $J_\xi\colon \M J_1\f E \to \Rn$. By Theorem~\ref{Th: 5.13}
$J_\xi$ \emph{is a conserved special function}. A momentum map $J$ is
\emph{equivariant} if for all $g\in G$ $J\circ\Phi_g=\CoAd_g\circ
J=J_{\Ad_{g^{-1}}}$, where $\Ad$ is the adjoint action on $\mathfrak{g}$ and
$\CoAd$ is the coadjoint action on $\mathfrak{g}^*$. There are several
obstructions for the existence of the momentum map in general. However, our
model always admits a momentum map on the domain of the Poincar\'e--Cartan form
$\Theta$.

\begin{Proposition}\label{sec:momentum-map-special}
  Let $\Phi$ be a projectable action of a Lie group $G$ of infinitesimal
  symmetries of the classical almost-cosymplectic-contact structure. Then, the
  map $J$ defined by $J_\xi=i_{\phi(\xi)}\Theta$ is an equivariant momentum
  map by conserved special functions for $\Phi$.
\end{Proposition}
\bpf
  It is easy to prove that $-d(i_{\phi(\xi)}\Theta)=i_{\phi(\xi)}\Ome$. This
  map is also equivariant: indeed,
  \begin{multline*}
    J_{\Ad_{g^{-1}}\xi}(j_1s(x))=\Theta(\Ad_{g^{-1}}\xi)(j_1s(x))=
    \Theta(T\Phi_{g^{-1}}\phi(\xi))(j_1s(x))=
    \\
    =\Phi^*_g\Theta(T\Phi_{g^{-1}}\phi(\xi))(j_1s(x))=
    J_{\xi}(\Phi_g(j_1s(x))).\QED
  \end{multline*}
\epf

\subsection{Examples}
Here we will briefly discuss momentum maps for symmetry groups acting on
Minkowski and Reissner--Nordstrom spacetimes.

\paragraph{Minkowski spacetime.} The Poincare group is acting on Minkowski
spacetime as its group of isometries. Let us denote by $\mathfrak{p}$ its Lie
algebra. After the introduction of adapted linear coordinates on $\f E$,
$\mathfrak{p}$ acts on $\f E$ in the standard affine way. As is well-known,
$\mathfrak{p}$ is the semi-direct product of $\Rn^4$ (representing
translations) and $\mathfrak{so}(1,3)$ (representing Lorentz transformations).
The momentum map takes the form
\begin{equation}
  \label{eq:9}
  J_\xi=i_{\phi(\xi)}(-\widehat\tau)=
  (mc/\hbar)\alpha^0(-\bar\phi(\xi)^0+x^1_0\bar\phi(\xi)^1
  +x^2_0\bar\phi(\xi)^2+x^3_0\bar\phi(\xi)^3),
\end{equation}
where $\bar\phi$ is the infinitesimal action on spacetime.
In particular, when $\xi=e_i\in\Rn^4$, then $\bar\phi(\xi)=\der_i$
and $J_\xi=(mc/\hbar)\alpha^0x^i_0$; when $\xi$ is a spacelike rotation
then $\bar\phi(\xi)=x^i\der_j-x^j\der_i$ and
$J_\xi=(mc/\hbar)\alpha^0(x^ix^j_0-x^jx^i_0)$.

\paragraph{Reissner--Nordstrom spacetime.} This spacetime is obtained by
requirements of spherical symmetry so its infinitesimal isometries are rotations
and, since the metric is static, time translations. The Lie algebra of the
corresponding Lie group is the semidirect product of $\Rn$ and
$\mathfrak{so}(3)$. The momentum map turns out to be
\begin{equation}
  \label{eq:11}
  J_\xi=i_{\phi(\xi)}(-\widehat\tau+\widehat{A}),
\end{equation}
where $\widehat A$ was given in \eqref{eq:13} and $\tau$ in~\eqref{eq:6}. Now,
when $\xi$ corresponds to time translation we have $\bar\phi(\xi)=\der/\der t$
and $J_\xi=(mc/\hbar)\alpha^0(-(1-k_s/r+k^2_q/r^2)-(q_0/(\hbar_0r)))$, which is
the Hamiltonian; when $\xi$ corresponds to spacelike rotations we convert the
vector field $\bar\phi(\xi)=x^i\der_j-x^j\der_i$ to spherical coordinates then
contract it with $-\widehat\tau+\widehat{A}$. We obtain a long coordinate
expression which is not difficult to compute and to which there is no
contribution from $\widehat A$.

\section{Perspectives}
\label{sec:perspectives}

We characterized infinitesimal phase symmetries of spacetime starting from a
quite natural Definition~\ref{df:phasesymm}. Various properties of such
symmetries have been derived, the most promising one coming from
Proposition~\ref{sec:momentum-map-special}. The Proposition opens the
possibility to reduce the almost-cosymplectic-contact structure along the lines
of \cite{Albert89,CC02,dLS93} (which are inspired by the classical
Marsden--Weinstein reduction).  Namely, since motions happen on level surfaces
of the form $J^{-1}(\mu)$ it is natural to restrict the dynamical system to
$J^{-1}(\mu)$. Then it could be proved that the resulting
almost-cosymplectic-contact structure would be degenerate along directions
tangent to group orbits in $J^{-1}(\mu)$. A quotient by the stabilizer of $\mu$
in $\mathfrak{g}^*$ should yield a new almost-cosymplectic-contact structure
for the reduced dynamical system. As far as we know, this reduction has never
been attempted in relativistic mechanics.

Another direction of investigation is given by enlarging the class of
symmetries under consideration. The requirement of projectability is quite
natural under the physical viewpoint, but it could be dropped in order to
consider symmetries of the generalized contact structure which depend on phase
space in an essentially non-projectable way, as we already mentioned in the
Introduction.  Noether symmetries of non-projectable type correspond to
conserved quantities that, in the standard $4$-velocity formalism, depend
polynomially on velocities \cite{San12}. The coefficients of monomials are
Killing tensors. This construction could be repeated in the framework of our
model. Killing-Yano tensors are also related with non-projectable symmetries
through a similar mechanism.

The task of our research has been the systematic exploration of infinitesimal
symmetries of the geometric structures on the general relativistic phase space.
The idea that we had in mind was to do a symmetry analysis in the framework of
generalized contact geometry. We pursued this task by characterizing
projectable infinitesimal symmetries of the generalized contact structure and
by relating such symmetries with a distinguished class of observables on the
phase space, the special phase functions. Moreover, we were able to relate
brackets of symmetries with brackets of special phase functions. Finally, we
proved that a momentum map for a group actions by projectable infinitesimal
symmetries always exist. Now, the most natural question that we can ask
ourselves is if it is possible to generalize the above results to the case of
non-projectable infinitesimal symmetries.  At the present moment, we cannot
predict which of the results in the paper could be extended to more general
symmetries. In particular, we do not know which class of conserved quantities
would correspond to non-projectable infinitesimal symmetries and if we would be
able to provide a bracket on such an unknown class; we also ignore if a
momentum map could be introduced for such symmetries. Maybe it would be
possible to introduce a distinguished class of non-projectable symmetries that
would be in correspondence with Killing and/or Killing-Yano tensors, and find
corresponding observables, a bracket between them etc..  The starting point
would be exactly relaxing the hypotheses of Theorem~\ref{Th: IS} in order to
allow for non-projectable symmetries.  This will be the subject of future work.

\end{document}